\documentclass[aps,prl,twocolumn,groupedaddress,superscriptaddress,amsfonts,amssymb,amsmath,citeautoscript,a4paper]{revtex4-1}

\usepackage{orcidlink}
\usepackage{lipsum}
\usepackage[utf8]{inputenc}
\usepackage[english]{babel}
\usepackage{microtype}
\usepackage{txfonts}
\usepackage{txfontsb}
\usepackage{bm}
\usepackage{xcolor}
\usepackage{graphicx}
\usepackage{float}
\usepackage{xspace}
\usepackage[skip=0pt, indent=15pt]{parskip}
\usepackage{titlesec}
\titleformat{\paragraph}[runin]{\normalfont\normalsize\bfseries}{}{0}{}[]
\titleformat{\section}[hang]{\normalfont\large\bfseries}{}{0}{}[]
\usepackage{subfiles}
\usepackage{enumerate}
\usepackage[inline]{enumitem}
\usepackage{multirow,rotating}

\usepackage{siunitx}
\DeclareSIUnit[number-unit-product=]\percent{\char`\%} 

\hypersetup{colorlinks,
            linkcolor={blue!75!black!80!yellow},
            citecolor={blue!75!black!80!yellow},
            urlcolor={blue!75!black!80!yellow},
            pdfstartview=FitH}

\makeatletter
\renewcommand{\fnum@figure}{FIG.~\thefigure}
\makeatother

\usepackage{xifthen}
\usepackage{etoolbox}
\usepackage{soul}

\newcommand{\todo}[1]{
	\textcolor{orange!80!yellow!95!black}{\textbf{[}%
	\ifthenelse{\isempty{#1}}%
		{\text{$\blacksquare$}}%
		{{\small\textsf{#1}}}%
	\textbf{]}}}

\usepackage{mdframed}
\definecolor{blue-violet}{rgb}{0.54, 0.17, 0.89}

\newmdenv[topline=false, rightline=false, bottomline=false,%
  linewidth=1.25pt, innerrightmargin=0pt, leftmargin=-8pt,%
  innerleftmargin=7pt, skipabove=0pt, skipbelow=0pt,%
  linecolor=blue-violet, fontcolor=blue-violet]{mdleftbar}

\begin{document}

\title{Prospects to bypass nonlocal phenomena in metals using phonon-polaritons}

\author{Jacob T. Heiden\,\orcidlink{0000-0003-4505-6107}}
\thanks{J.~T.~H. and E.~J.~C.~D. contributed equally to this work.}
\affiliation{School of Electrical Engineering, Korea Advanced Institute of Science and Technology (KAIST), Daejeon 34141, Korea}

\author{Eduardo~J.~C.~Dias\,\orcidlink{0000-0002-6347-5631}}
\thanks{J.~T.~H. and E.~J.~C.~D. contributed equally to this work.}
\affiliation{POLIMA---Center for Polariton-driven Light--Matter Interactions, University of Southern Denmark, Campusvej 55, DK-5230 Odense M, Denmark}

\author{Minhyuk~Kim}
\address{Graduate School of Semiconductor Materials and Devices Engineering, Ulsan National Institute of Science and Technology, Ulsan, Korea}

\author{Martin~Nørgaard\,\orcidlink{0009-0002-5412-4669}}
\affiliation{POLIMA---Center for Polariton-driven Light--Matter Interactions, University of Southern Denmark, Campusvej 55, DK-5230 Odense M, Denmark}

\author{Vladimir~A.~Zenin\,\orcidlink{0000-0001-5512-8288}}
\affiliation{Center for Nano Optics, University of Southern Denmark, Campusvej 55, DK-5230~Odense~M, Denmark}

\author{Sergey~G.~Menabde\,\orcidlink{0000-0001-9188-8719}}
\affiliation{School of Electrical Engineering, Korea Advanced Institute of Science and Technology (KAIST), Daejeon 34141, Korea}

\author{Hu~Young~Jeong\,\orcidlink{0000-0002-5550-5298}}
\address{Graduate School of Semiconductor Materials and Devices Engineering, Ulsan National Institute of Science and Technology, Ulsan, Korea}

\author{N.~Asger~Mortensen\,\orcidlink{0000-0001-7936-6264}}
\thanks{\href{mailto:asger@mailaps.org}{asger@mailaps.org}; \href{mailto:jang.minseok@kaist.ac.kr}{jang.minseok@kaist.ac.kr};}
\affiliation{POLIMA---Center for Polariton-driven Light--Matter Interactions, University of Southern Denmark, Campusvej 55, DK-5230 Odense M, Denmark}
\affiliation{Danish Institute for Advanced Study, University of Southern Denmark, Campusvej 55, DK-5230 Odense M, Denmark}

\author{Min~Seok~Jang\,\orcidlink{0000-0002-5683-1925}}
\thanks{\href{mailto:asger@mailaps.org}{asger@mailaps.org}; \href{mailto:jang.minseok@kaist.ac.kr}{jang.minseok@kaist.ac.kr};}
\affiliation{School of Electrical Engineering, Korea Advanced Institute of Science and Technology (KAIST), Daejeon 34141, Korea}
\email{jang.minseok@kaist.ac.kr}

\date{\today}

\begin{abstract}
Electromagnetic design relies on an accurate understanding of light-matter interactions, yet often overlooks electronic length scales. Under extreme confinement, this omission can lead to nonclassical effects, such as nonlocal response. Here, we use mid-infrared phonon-polaritons in hexagonal boron nitride (hBN) screened by monocrystalline gold flakes to push the limits of nanolight confinement unobstructed by nonlocal phenomena, even when the polariton phase velocity approaches the Fermi velocities of electrons in gold. We employ near-field imaging to probe polaritons in nanometre-thin crystals of hBN on gold and extract their complex propagation constant, observing effective indices exceeding 90. We further show the importance of sample characterisation by revealing a thin low-index interfacial layer naturally forming on monocrystalline gold. Our experiments address a fundamental limitation posed by nonlocal effects in van der Waals heterostructures and outline a pathway to bypass their impact in high-confinement regimes.
\end{abstract}

\maketitle


\section{Introduction}
Nanophotonic studies increasingly rely on mesoscopic features and corresponding highly confined electromagnetic fields, with two-dimensional (2D) van~der~Waals (vdW) crystals, such as graphene and hexagonal boron nitride (hBN), playing pivotal roles~\cite{Basov:2016,Moon:2023}. The polar nature of hBN facilitates the hybridisation of optical phonons (lattice vibrations) with photons to form collective light-matter modes called phonon-polaritons (PhPs). Similarly, graphene’s semi-metal nature allows coupling between the collective oscillations of its free charge carriers and photons, resulting in highly localised plasmonic modes. These phenomena provide powerful means to localise and control electromagnetic fields in mid-infrared (MIR) to terahertz frequencies~\cite{Basov:2021}.

In mesoscopic plasmonics, it has become evident that the ultra-high field confinement leads to considerable discrepancies between classical electromagnetic predictions and experimental observations~\cite{Ciraci:2012,Raza:2015b,Yang:2019,Mortensen:2021,Boroviks:2022}. These discrepancies arise from the neglect of quantum-mechanical effects, which mainly become significant when electrons are confined in low-dimensional systems. Typically, individual aspects of these effects, such as spill-out~\cite{Zhu:2016} and nonlocality~\cite{Raza:2015a}, can be addressed using phenomenological approaches~\cite{Mortensen:2014} or the unified framework of the Feibelman $d$-parameters~\cite{Feibelman:1982,Yan:2015,Christensen:2017,Yang:2019,Mortensen:2021}. Regardless of the approach, there is a strong desire to probe these nonclassical effects further and to deepen the understanding of how to support or circumvent them in state-of-the-art electromagnetic designs. 

In planar structures with in-plane translational invariance, elastic scattering among electromagnetic and matter excitations conserves energy ($\hbar\omega$, where $\omega$ is the angular frequency) and in-plane momentum ($q$), implying that this requires modes with the same phase velocity ($v_\phi=\omega/q$) as a prerequisite. In the MIR and terahertz regimes, acoustic-like graphene plasmons (AGPs) -- graphene plasmons screened by a nearby metallic substrate such as gold (Au)~\cite{Alonso-Gonzalez:2017,Menabde:2021} -- have been used to confine light into nanometre-scale spacers, reducing the plasmon phase velocity $v_\phi$ to approach the Fermi velocity of graphene $v_{F,\mathrm{G}}\simeq 10^6$\,m/s in an attempt to characterise the quantum response from graphene~\cite{Lundeberg:2017}. AGPs have been proposed as a potential probe for quantum-corrected electrodynamics associated with nearby metals~\cite{Goncalves:2020,Goncalves:2021} and even excitations in more strongly-correlated matter~\cite{Berkowitz:2021,Costa:2021}. However, the potentially significant nonlocal effects in the adjacent metal cannot be ignored \emph{a priori}, as the Fermi velocity of the commonly used Au, $v_{F,\mathrm{Au}}\simeq 1.4\times 10^6$\,m/s, is comparable to or even greater than that of graphene~\cite{Dias:2018,Goncalves:2020,Goncalves:2021}. 

As an alternative to disentangle the nonlocal effects in graphene and metal~\cite{Goncalves:2021}, it may be rewarding to remove graphene from the system altogether and employ a different polariton material instead. Polar vdW crystals, such as hBN, support hyperbolic PhPs, whose dispersion can be modified by the material geometry, surrounding environment, and crystal thickness~\cite{Dai:2014,Caldwell:2014}. Notably, experiments have shown that PhPs are supported by a single atomic layer of hBN, achieving an effective mode index ($n_{\mathrm{eff}}=q/q_0$ where $q_0=\omega/c$ is the free-space wavevector) approaching 60~\cite{Dai:2019}. However, PhP confinement can be further increased by placing hBN on a metallic substrate, where the polaritonic fields are screened by the image charges in the metal, forming a hyperbolic image phonon-polariton (HIP) mode~\cite{Ambrosio:2018,Lee:2020,Menabde:2022b}, similar to the screening mechanism giving rise to AGPs~\cite{Menabde:2022a}. Compared to the PhP in freestanding hBN, the HIP mode exhibits approximately four-fold lateral confinement, with the strongest field enhancement occurring at the interface between hBN and metal~\cite{Menabde:2022a} -- in this work, we use crystalline Au.

\begin{figure}[tb]
\centering
\includegraphics[width=\columnwidth]{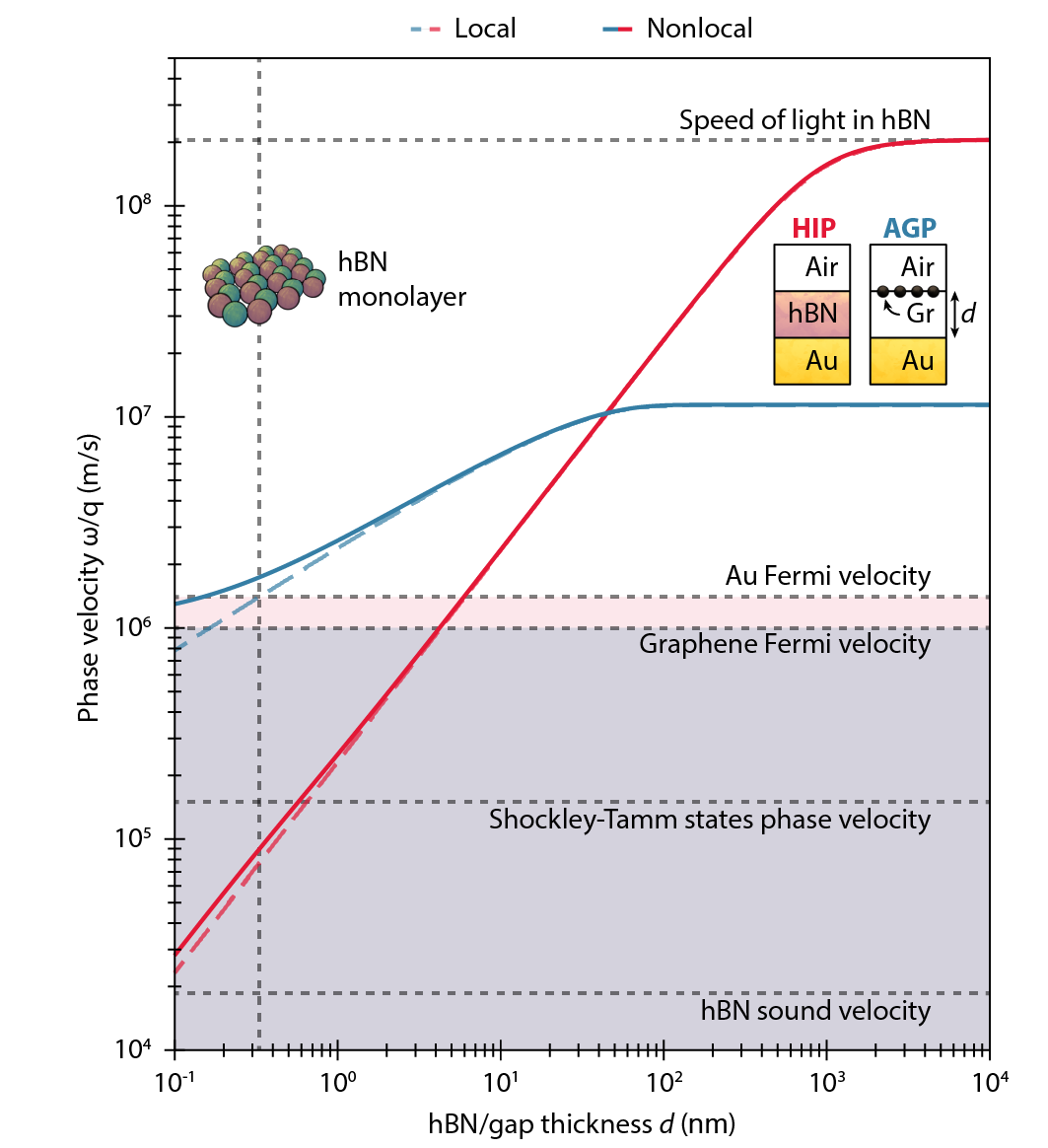}
\caption
{\textbf{Prospects for extreme confinement and slow-down of HIP modes.} Effect of hBN thickness on HIP phase velocity (red lines) and spacer thickness on AGP phase velocity (blue lines) at a frequency of 1500\,cm$^{-1}$. Graphene is modelled with Fermi level $E_F=0.5$\,eV and damping $\hbar\gamma=10$\,meV. Horizontal grey dotted lines indicate possible nonlocal couplings to slow matter excitations. For AGPs, the phase velocity is influenced by the combined Fermi velocities of graphene and Au, leading to significant discrepancies between local and nonlocal theories -- the phase velocity cannot bypass the Fermi velocity barrier. For HIPs, the Fermi velocity of Au does not induce notable deviations, and the mode can be slowed beyond the intuitive phase velocity barrier. At an hBN thickness of 1\,nm (corresponding to three atomic layers), the HIP phase velocity is approaching the phase velocity of plasmons associated with Shockley--Tamm surface states in Au yet is still far from the sound velocities causing spatial dispersion in hBN. At large thicknesses, the phase velocity is naturally bounded by the speed of light in the medium.}
\label{fig1}
\end{figure}

We note that both hBN and crystalline Au potentially introduce their own spatially dispersive response, being associated with phonons~\cite{Gubbin:2020} and Shockley--Tamm surface states~\cite{RodriguezEcharri:2021}, respectively (Fig.~\ref{fig1}). However, for the hBN, this response only manifests at very large wave vectors due to the relatively slow phonons with sound velocities being anisotropic and in the range of $3\times 10^3$\,m/s to $2\times 10^4$\,m/s~\cite{Jimenez-Rioboo:2018}. Likewise for the Au, the AGP-like plasmons associated with the Shockley--Tamm surface states have a phase velocity of $1.5\times 10^{5}$\,m/s~\cite{RodriguezEcharri:2021}. Thus, the Fermi velocity of Au, being the highest among all other characteristic velocities in the hBN/Au system, intuitively prompts two pivotal and entwined questions: what are the prospects for probing nonlocal phenomena in metals using phonon-polaritons, and can they offer a pathway to bypass the perceived Fermi velocity barrier?

Here, we use thin crystals of naturally abundant hBN on monocrystalline Au flakes to probe HIPs with $n_{\mathrm{eff}}>90$, where the measured oscillation periodicity is $\lambda_{\mathrm{HIP}}/2\simeq35$\,nm -- close to the spatial resolution limit of scattering-type scanning near-field optical microscopy (s-SNOM) used here -- in $\approx11$\,nm-thick hBN. In our experiments, the atomically flat surface of the Au is critical to dramatically suppress roughness-induced phonon-polariton scattering and accurately extract the complex propagation constant of the ultra-confined HIPs~\cite{Menabde:2022b}. We observe a HIP dispersion deviation of up to 30\% between the classical predictions and the experiments, far exceeding what nonlocal effects alone can explain. However, this discrepancy can ultimately be attributed to a carbonous layer between hBN and Au, likely native to Au. This ultrathin layer fully accounts for the observed dispersion perturbations, demonstrating that nonlocal effects are minuscule in the system at the probed polariton velocities near $v_{F,\mathrm{Au}}$. While AGPs are known to exhibit strong nonlocality-driven saturation of the phase velocity when pushed to high confinement, HIPs behave purely local, clearing the path towards record-breaking field confinement in phonon-polaritons (Fig.~\ref{fig1}).

\section{Results}
\begin{figure*}[ht]
\centering
\includegraphics[width=\textwidth]{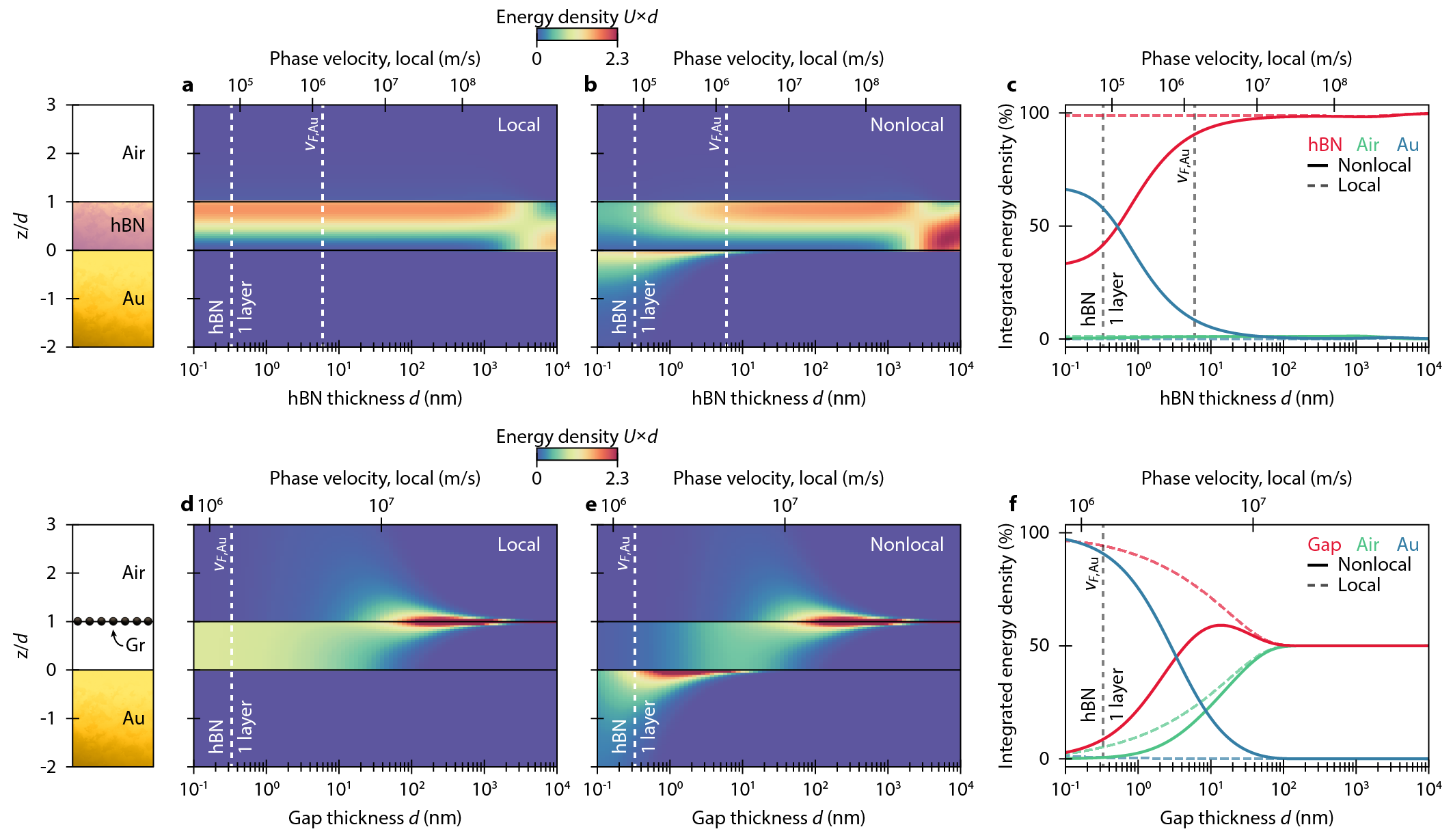}
\caption
{\textbf{Energy density distribution.} \textbf{a},\textbf{b}, Local (\textbf{a}) and nonlocal (\textbf{b}) energy density $U$ (multiplied by the spacing distance $d$) as a function of $d$ and position along the HIP-supporting heterostructure (schematic on the left), at a frequency of 1500\,cm$^{-1}$. Energy density is calculated and normalised as described in the Supplementary information. \textbf{c}, Integrated energy density within the hBN (red), air (green), and Au (blue) regions, as percentage of the total integrated energy density in the full structure, plotted as a function of $d$. Results are calculated within both local (dashed) and nonlocal (solid) metal formalisms. \textbf{d},\textbf{e},\textbf{f}, same as (\textbf{a}), (\textbf{b}), (\textbf{c}), respectively, but for the AGP-supporting heterostructure depicted on the left. Graphene conductivity is always described through its nonlocal response. In all panels, dashed lines are added highlighting the hBN monolayer limit and the Fermi velocity of Au.}
\label{fig2}
\end{figure*}

\begin{figure*}[ht]
\centering
\includegraphics[width=\textwidth]{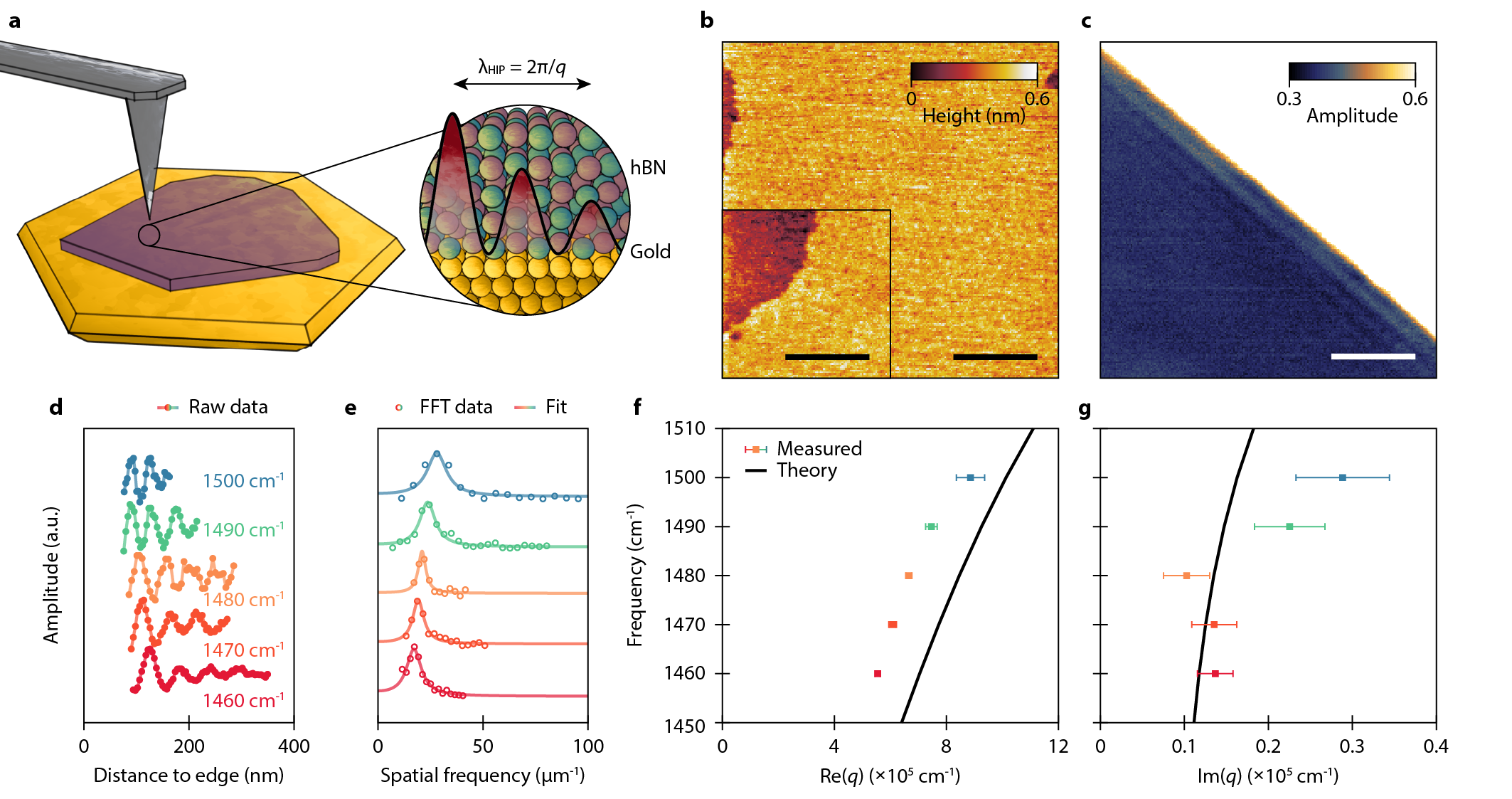}
\caption
{\textbf{Experimental setup and near-field imaging.} \textbf{a}, Experimental heterostructure: monocrystalline Au flake with an hBN flake on top, probed by a metallic AFM tip. Inset to \textbf{a}, close-up of the hBN/Au interface displaying interference fringes probed by s-SNOM with a in-plane wave vector $q=2\pi/\lambda_\mathrm{HIP}$ \textbf{b}, AFM measurement revealing terraces on the Au surface. Inset to \textbf{b}, surface terrace in the top left corner of the main image (scale bars, 250\,nm). \textbf{c}, Spatial plot of the s-SNOM measurement of a sample at a frequency of 1480\,cm$^{-1}$, where the field is saturated over the Au area. (scale bar, 250\,nm). \textbf{d}, Line scans extracted from \textbf{c} and at frequency increments of 10\,cm$^{-1}$ between 1460 and 1500\,cm$^{-1}$. The line scans were all normalised and offset for ease of comparison. \textbf{e}, Normalised and offset Fourier spectrum and fit of the tip-launched part of the near-field signal in \textbf{d}. \textbf{f},\textbf{g}, Experimental and theoretical dispersion relation, $q_0$ versus $\mathrm{Re}(q)$ (\textbf{f}) and $\mathrm{Im}(q)$ (\textbf{g}).}
\label{fig3}
\end{figure*}

We start with a theoretical investigation of HIPs supported by a thin hBN layer on Au and AGPs supported by a graphene sheet with Fermi level $E_F=0.5$\,eV, suspended a small distance above Au. At a frequency of 1500\,cm$^{-1}$ our calculations predict that even in about 5--10\,nm-thick hBN flakes, HIPs can be slowed down to phase velocities near $v_{F,\mathrm{Au}}$ in the middle of hBN's upper Reststrahlen band, whereas AGPs require Ångström-scale spacers to reach similar phase velocities (Fig.~\ref{fig1}). Moreover, our theoretical implementation of Au nonlocality (see Methods and Supplementary information section S1) suggests that HIPs can, in principle, bypass the perceived $v_{F,\mathrm{Au}}$ barrier, reaching phase velocities orders of magnitude slower than supported by AGPs.

Why, then, do HIPs theoretically behave largely in a classical manner and seemingly independent of the Au Fermi velocity? To get a better grasp on this, we study the impact of metal nonlocal effects on the energy density distribution $U$. Figure~\ref{fig2}a shows that, in the local case, the energy density is almost entirely concentrated within the hBN and remains invariant with respect to $d$, except at $d>1000$\,nm, where the mode becomes fundamentally limited by the speed of light (Fig.~\ref{fig1}). Interestingly, considering nonlocal effects, echo this overall trend while providing a more complete picture of the metal's screening behaviour (Fig.~\ref{fig2}b): As the mode approaches $v_{F,\mathrm{Au}}$ at $d\simeq10$\,nm, the Au still effectively screens the PhPs, as expected of a good metal, and continues to do so, to a reasonable extent, at smaller thicknesses. Only at $d\lesssim2$\,nm -- where the HIP phase velocity is nearly an order of magnitude lower than $v_{F,\mathrm{Au}}$ -- does the energy density in the Au become visually significant, leading to a noticeable redistribution from the hBN.

To quantify this electromagnetic energy redistribution, we integrate $U$ along the $z$-axis (Fig.~\ref{fig2}c), confirming that the HIP mode stays highly confined in the hBN, with Au contributing less than 10\% until $d\simeq5$\,nm, and becoming dominant only for $d\lesssim0.5$\,nm. Notably, as the thickness decreases further, nonlocal effects appear to saturate, suggesting that the mode volume can, in theory, be reduced indefinitely and in practice it is limited only by the intrinsic atomic thickness of monolayer hBN.

In contrast to the energy density distribution of HIP modes, small-gap AGPs exhibit markedly different characteristics. In the local calculations, the AGP mode emerges at $d\lesssim100$\,nm and remains increasingly within the gap between graphene and Au (Fig.~\ref{fig2}d). However, nonlocal effect significantly alter this picture: almost immediately after the mode formation, energy density penetration into Au is apparent, with a much higher amplitude than for the HIP mode (Fig.~\ref{fig2}e). Integrating $U$ along $z$ reiterates the early metal penetration, as the AGP surpass 10\% Au contribution at $d\simeq16$\,nm and the contribution becomes dominant at $d\lesssim3.5$\,nm (Fig.~\ref{fig2}f). This behaviour is striking when contrasted with the HIP mode, demonstrating that the AGP mode exhibit nonclassical behaviour well before reaching $v_{F,\mathrm{Au}}$ while HIPs do not.

Importantly, the spatial distribution of energy density distinguishes these modes: while the HIP mode remains concentrated near the hBN/air interface (Fig.~\ref{fig2}b), the AGP mode exhibits a more uniform distribution within the gap (Fig.~\ref{fig2}e). We speculate that this distribution in the HIP mode reduces energy penetration into the metal, thereby mitigating the influence of $v_{F,\mathrm{Au}}$.

Beyond the main text, we provide a deeper dive into the theoretical aspects of nonlocal effects with additional figures in Supplementary information section~S2 detailing: field profiles (Fig.~S1); additional comparisons between HIP and AGP modes (Figs.~S2 and S3); and spectral variations of the plots figure~\ref{fig2}c,f (Fig.~S4). Interestingly, the HIP mode exhibits a negative dispersive behaviour concerning its small nonlocal effects, where an increase in $q$ -- driven by a frequency increase -- results in a smaller perturbation. Additionally, we compare the nonlocal effects in Au and titanium (Ti) substrates (see Supplementary information~S3), as Ti is known to exhibit a stronger nonlocal response in AGPs~\cite{Dias:2018}, which we also observe in HIPs.

Our experimental approach for measuring highly confined HIP modes employs s-SNOM, which enables direct probing of polaritonic fields (Fig.~\ref{fig3}a). Operating within the upper Reststrahlen band of hBN, in the MIR range, we utilise a laser frequency spanning from 1460 to 1510\,cm$^{-1}$, although for some samples, poor signal-to-noise ratio prevents high-frequency imaging. When the metallic s-SNOM tip approaches the edge of an hBN flake, the near-field interference fringes emerge from the superposition of the tip-launched and edge-reflected HIP. The periodicity of the observed oscillations reveals the HIP wavelength as it is equal to $\lambda_\mathrm{HIP}/2$, from which the in-plane wavevector $q = 2\pi/\lambda_\mathrm{HIP}$ can be derived (Fig.~\ref{fig3}a, close-up inset). By using thin ($t<40$\,nm) hBN flakes, the strong confinement and significantly reduced phase velocity of the HIPs provide an opportunity to test whether nonlocal effects associated with the metal's Fermi velocity constrain the ability to slow down the phonon-polaritons below this critical velocity.

Additionally, to ensure precise HIP characterisation, we measure the surface roughness of the Au flakes prior to hBN deposition (Fig.~\ref{fig3}b), revealing an ultrasmooth surface with an RMS roughness of $\approx 44$\,pm, along with atomic terraces indicative of pristine surfaces. Atomically smooth surfaces are a requisite for reducing noise in s-SNOM measurements of nanoscale features, especially for fundamental image modes where the $E_z(z)$ field is strongest at the interfaces (see Supplementary information S2). In contrast, evaporated and sputtered Au surfaces -- where attaining RMS roughnesses on the order of 0.5\,nm typically represents the best case~\cite{Yang:2019,Menabde:2022b} -- are known to be insufficient for precise near-field investigation of HIP modes, even when the hBN thickness exceeds 20\,nm, leading to a noisy signal and increased loss through surface-mediated scattering~\cite{Menabde:2022b}.

Using the s-SNOM method, we spatially map the HIP-associated interference fringes (Fig.~\ref{fig3}c) and extract the near-field profile by performing a line scan perpendicular to the hBN edge (Fig.~\ref{fig3}d). Ideally, the hBN edge is straight, as it ensures a uniform wavefront and allows for averaging across multiple parallel lines, thereby improving the signal-to-noise ratio. For accuracy, it is beneficial to truncate the signal at the second fringe from the hBN crystal edge, as the first fringe is contaminated by parasitic near-field signals resulting from strong edge scattering. The shortest observed fringe is $\approx35$\,nm at 1500\,cm$^{-1}$, which is comparable to the spatial resolution limit of s-SNOM. Although higher resolution and better coupling between the s-SNOM tip and HIP can be achieved by using sharper tips, this comes at the cost of reduced signal-to-noise ratio, as the s-SNOM signal strongly depends on tip radius, ultimately creating a trade-off between resolution and contrast~\cite{Fei:2011}. Light-matter coupling is further constrained by the intrinsic Reststrahlen band of hBN, limiting the near-field signal contrast~\cite{Paarmann:2024} and by the decreasing field penetration depth into the free-space above thinner hBN flakes~\cite{Menabde:2021,Menabde:2022b} (see Supplementary information S4).

We note that another periodic component of the near-field signal may arise from interference between the hBN-edge-launched mode and the illumination beam, having an oscillation periodicity of $\lambda_\mathrm{HIP}$. However, this signal is very weak in thin hBN flakes, generally limiting its impact on our measurements. Due to the increased fringe periodicity, this edge-excited signal is conceptually desirable, and it can be enhanced by positioning the hBN flake over the edge of an Au substrate~\cite{Menabde:2022b}, or by patterning a nanoantenna on top of the hBN. However, these experimental configurations introduce additional complexities during characterisation; specifically, they introduces a $q$-dependency on the laser’s incidence angle in s-SNOM~\cite{Huber:2005,Wong:2021,Casses:2022,Jang:2024,Casses:2024}, as well as the potential for an air gap between hBN and Au in the hBN-over-Au-edge case, or sample damage and contamination during nanoantenna fabrication. For these reasons, we avoid the edge-excitation approach here.

Extensive research has been done on the methodology of s-SNOM analysis~\cite{Casses:2022,Jang:2024,Casses:2024}; therefore, we summarise our characterisation process here, while outlining it in detail in Supplementary information S5. To precisely extract $q$, we perform a fast Fourier transform (FFT) on each line scan and fit the resulting spectrum with that of a damped harmonic oscillator model. The peak in the Fourier spectrum directly provides the spatial frequency, from which $q$ is obtained (Fig.~\ref{fig3}e). In order to determine the loss of the HIP mode, we apply an inverse transform to the Fourier signal, stemming from the tip-launched mode, and fit it to a damped harmonic oscillator, incorporating the $x^{-1/2}$ geometrical decay factor from radial-wave spreading~\cite{Jang:2024}. We also employ this approach to establish the dielectric function of our hBN crystals by conducting experiments on hBN of varying thickness on a silicon (Si) substrate ahead of examining the hBN/Au heterostructure (Supplementary information section S6).

Figure~\ref{fig3}f displays the $\mathrm{Re}(q)$ we extract from the measurements, showing a near-linear dispersion that is a characteristic of image modes at low frequencies. To the best of our knowledge, the measured $\mathrm{Re}(q) = (8.86 \pm 0.51) \times 10^5\,\mathrm{cm}^{-1}$ and the associated $n_\mathrm{eff}=94$ represent the highest values observed for hBN in near-field studies, even exceeding those in a single atomic layer of hBN measured at lower frequencies~\cite{Dai:2019}. Since the near-field contrast diminishes for thinner vdW crystals, many studies have adopted a method of measuring only a single fringe to extract $\mathrm{Re}(q)$, enabling the probing of highly confined polaritons~\cite{Dai:2014,Zhi:2015,Dai:2019}. While this single-fringe approach has been pivotal in advancing high-confinement near-field studies, the analysis of multiple fringes, as we achieve here, provides greater accuracy.

Analysis of multiple near-field fringes allows us to extract the imaginary part of the HIP's wavevector, $\mathrm{Im}(q)$, shown in Figure~\ref{fig3}g. The observed $\mathrm{Im}(q)$ values are substantial; however, this result does not necessarily indicate the presence of nonclassical effects such as surface-enabled Landau damping. The high loss is primarily because $\mathrm{Im}(q)$ is not independent of $\mathrm{Re}(q)$. Moreover, these specific measurements for the most part closely resemble classical predictions, leaving any deviations ambiguous and challenging to interpret definitively. A detailed characterisation, involving multiple samples with varying hBN thickness, which we provide later, demonstrates a consistent trend of the blueshift for $\mathrm{Re}(q)$ and $\mathrm{Im}(q)$ -- contrary to what would be expected from nonlocal effects and Landau damping. 

Despite the greater accuracy, our measurements reveal a significant blueshift relative to the classical predictions of $\mathrm{Re}(q)$, with deviations exceeding 20\%. Such a significant deviation from classical theory is rarely observed in experiments. For example, state-of-the-art gap surface plasmon investigations require a spacer of approximately 1\,nm to observe such deviations~\cite{Yang:2019}, a behaviour echoed by AGPs~\cite{Iranzo:2018}, suggesting the need for an alternative explanation. 

\begin{figure}[tb]
\centering
\includegraphics[width=\columnwidth]{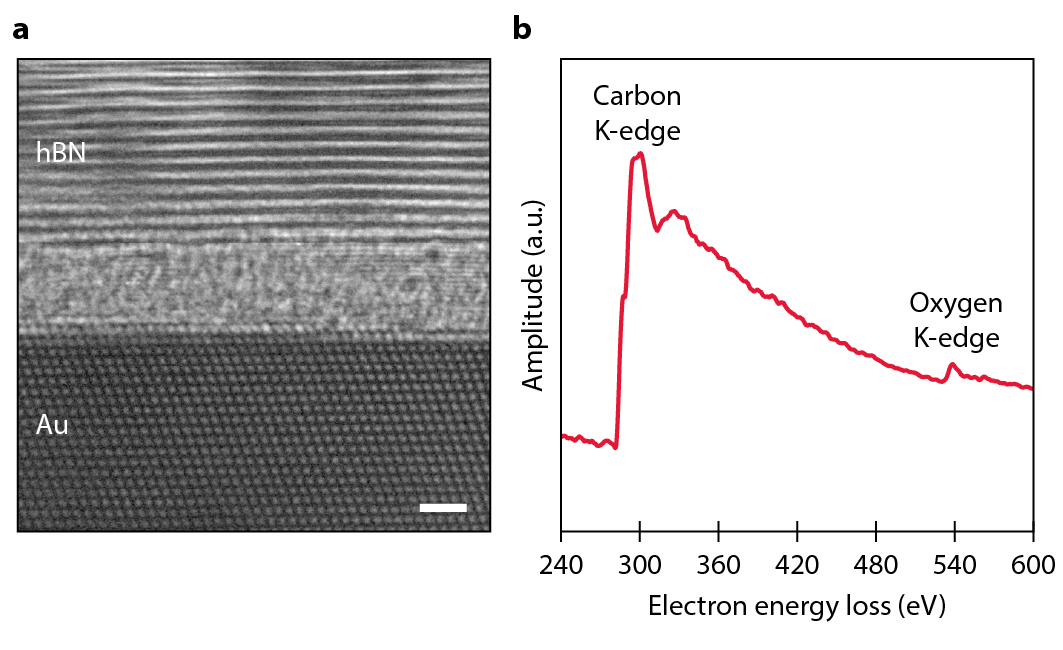}
\caption
{\textbf{Observation of interfacial layer.} \textbf{a}, cross-sectional high-resolution TEM image of the hBN/Au interface where an interfacial layer is present (scale bar, 1\,nm). \textbf{b}, EELS spectra of the interfacial layer with clear peaks at the carbon and oxygen K-edges.}
\label{fig4}
\end{figure}

To that end, we capture high-resolution TEM images of several samples, focusing on the interface between Au and hBN (Fig.~\ref{fig4}a). These images confirm the crystalline orientation of our Au crystals with visible atomic terraces of approximately 0.22\,nm height. The expected fringes characteristic of crystalline hBN, with an atomic spacing of around 0.33\,nm, are also clearly observed. However, between the two materials we detect an amorphous interfacial layer, typically ranging from 1.3 to 2\,nm in thickness. Notably, annular dark-field scanning TEM, which is frequently used to characterise cross-sections, would conceal the presence of an interfacial layer, as the weak electron scattering of dielectrics makes it difficult to unambiguously distinguish it from hBN (see Supplementary information S7). This interlayer could account for the discrepancies between classical predictions and experimental data, depending on its optical properties -- particularly, a lower refractive index would lead to greater deviations of $q$ and especially $\mathrm{Re}(q)$, as it reduces the effective index of the HIP modes.

\begin{figure*}[htb]
\centering
\includegraphics[width=\textwidth]{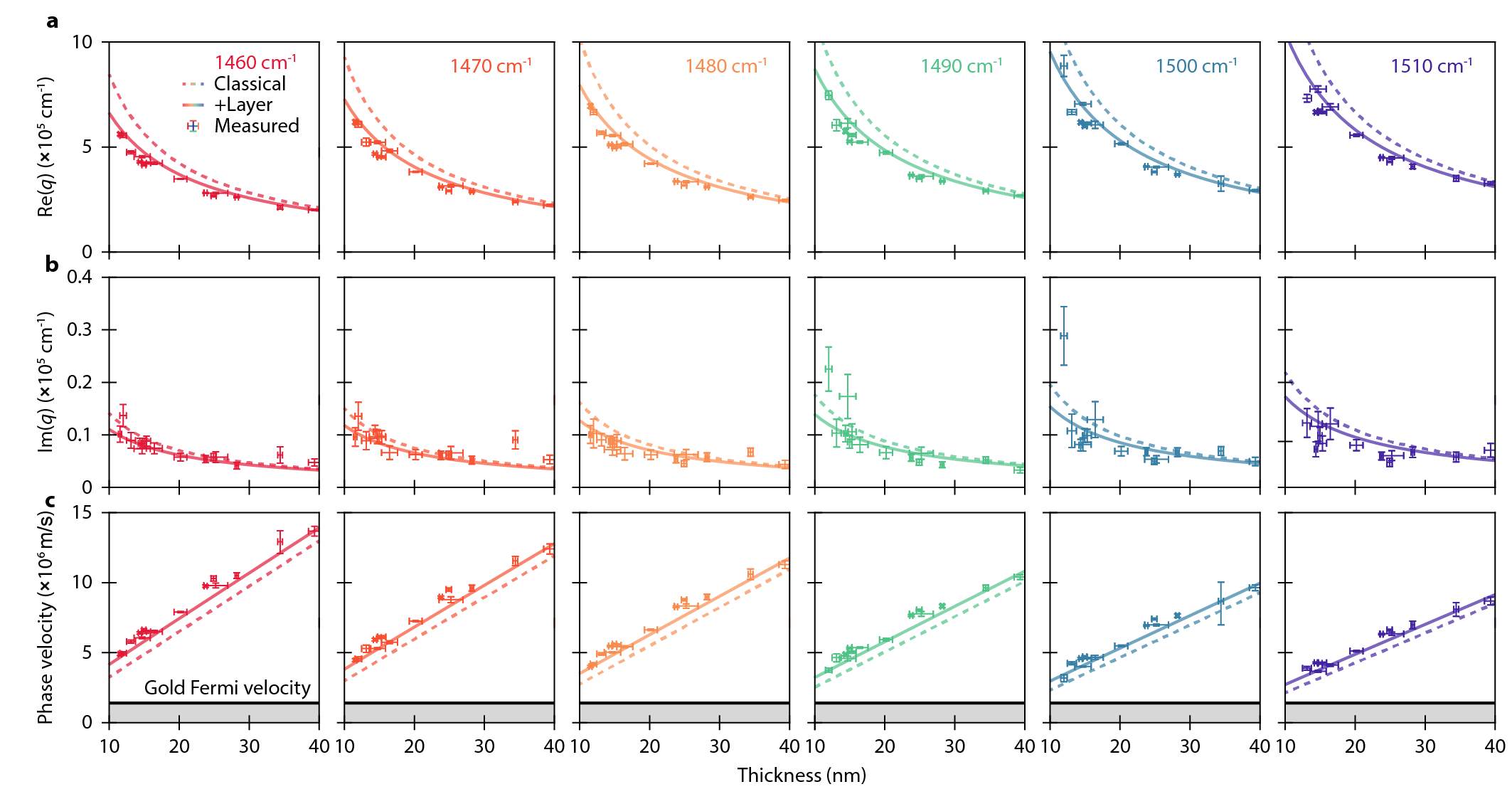}
\caption
{\textbf{Systematic measurement of the dispersion.} \textbf{a},\textbf{b},\textbf{c}, Measured (markers) dispersion of $\mathrm{Re}(q)$ (\textbf{a}), $\mathrm{Im}(q)$ (\textbf{b}), and the phase velocity (\textbf{c}) as a function of hBN thickness and HIP frequency. Classical theory (dashed lines) deviate considerably from our observations but are well accounted for when adding an interfacial layer (solid lines) in the classical theory.}
\label{fig5}
\end{figure*}

To understand the composition of this interfacial layer, we perform electron energy loss spectroscopy (EELS) using TEM. The analysis reveals that the layer is composed primarily of carbon and oxygen (Fig.~\ref{fig4}b), suggesting that it optically behaves like an organic material. Although the formation of surface layers on Au is rarely considered in nano-optics, it is in fact a well-established phenomenon~\cite{Smith:1980,Turetta:2021}. Surface layers, predominantly composed of carbon, but also oxygen, begin to form within minutes on clean Au surfaces~\cite{Turetta:2021}. While we attempt to mitigate such layers by cleaning our Au crystals with oxygen plasma (Methods), which temporarily leaves a clean surface, the plasma-treated surface may expedite the formation of a new layer before the hBN deposition can be completed -- we observe the formation of these layers as well (see supplementary information S8). Thus, avoiding the formation of this layer entirely would likely require operating in a controlled environment, such as inert nitrogen or argon chambers, or a high vacuum environment throughout the entire process, or employing the alternative fabrication methods such as Au deposition on hBN. However, while thermal- and sputter-deposited Au can achieve direct contact, they likely introduce hBN lattice defects as observed for MoS$_2$~\cite{Hong:2023} -- which is also a reason why we avoid nanoantenna fabrication on our samples.

Given the likely organic nature of the interfacial layer, we estimate its optical constants by referencing to previously reported values for organic compounds. Across the mid-infrared frequency range of our study, organic solids and liquids typically exhibit refractive indices between 1.35 and 1.5, with some, such as methyl groups and their esters~\cite{Myers:2017}, and polydimethylsiloxane (PDMS)~\cite{Zhang:2020}, approaching values just under 1.3. At the same time, extinction coefficients for these materials are generally small, on the order of $10^{-2}$. Using these parameters, we include the lossless interfacial layer into the model by assuming a thickness of 2\,nm and a real refractive index of 1.2.

Figure~\ref{fig5} summarises our s-SNOM results by comparing $q$ across multiple samples with varying hBN thickness, with the aim of identifying any systematic trend in the discrepancies. To this end, we contrast our experimental results with predictions from classical electrodynamic theory with and without incorporating the interfacial layer. Our analysis reveals that the HIP mode is consistently blueshifted relative to the classical prediction by up to as much as 30\% as the flake thickness decreases, but the inclusion of the interfacial layer accounts for this perturbation (Fig.~\ref{fig5}a). However, we note that for the hBN flakes around and above 30\,nm, the inclusion of an interfacial layer is not particularly relevant as the deviation of $\mathrm{Re}(q)$, across all frequencies, is negligible.

When analysing $\mathrm{Im}(q)$, as a function of thickness (Fig.~\ref{fig5}b), we find a similar trend where the experiment yields lower values than the theory. However, the measured loss has a higher degree of data variability and the sporadic presence of outliers, leading to increased uncertainties. This is anticipated due to the reduced near-field contrast associated with highly confined HIPs, as discussed above (additionally, see Supplementary information S4). Despite the relatively noisy data, the inclusion of the interfacial layer well accounts for the observed deviations, signifying no additional damping. Similar observations have been reported for AGPs~\cite{Iranzo:2018}, attributed to Landau damping dissipating for $\omega \ll \omega_p$, where $\omega_p$ is the metal plasma frequency. This phenomena is well explained by the imaginary part of the Feibelman $d$-parameters approaching zero at low frequencies~\cite{Yang:2019,Goncalves:2021}.

Finally, to quantify the measurements and their relation to classical electrodynamic theory, we calculate the HIP phase velocity from $\mathrm{Re}(q)$ in figure~\ref{fig5}a. As the classical theory predicts, a linear relationship is evident between the hBN thickness and the phase velocity (Fig.~\ref{fig5}c). A consistent linear deviation of the experimental data from theory is effectively accounted for by including the interfacial layer in the model. Notably, our measurements show no evidence of deviations from classical electrodynamics as we approach the Fermi velocity of Au, even when probing phase velocities as low as 
$\approx 2v_{F,\mathrm{Au}}$. We would like to emphasise that such phase velocities are comparable to those demonstrated for AGPs with an atomically thin hBN spacer where significant nonlocal effects have been observed, signalling the limiting field confinement~\cite{Iranzo:2018}. In contrast, our experimental results and theoretical calculations suggest that nonclassical effects are not exhibited for HIP modes in the hBN/Au heterostructure even when their phase velocity goes below the Fermi velocity of Au, paving the way towards uncharted territories of field confinement.

\section{Discussion}
The significant advancements in nanophotonics initiated by plasmons in metals and graphene have a fundamental electrodynamic constrain: at mesoscopic confinements, plasmons exhibit pronounced nonclassical effects. In contrast, HIPs in hBN offer an  alternative, achieving high field confinement with a substantially lower nonlocality to wavevector ratio (see Supplementary information S2). This ability to effectively mitigate nonclassical effects presents opportunities in a plethora of fields, including spontaneous emission control, quantum emitter coupling, and the exploration of velocity-dependent phenomena such as Shockley--Tamm surface states and phonon nonlocality~\cite{Monticone:2025}. 

Although our experiments show that nonclassical corrections are unnecessary in our hBN-Au setup, with nonlocal theory predicting a perturbation of less than 1\% for a 10\,nm thick hBN slab, the Fermi velocity remains a crucial factor in nonlocal response considerations for electromagnetic design. For example, replacing Au with Ti -- a widely used adhesion layer -- leads to an order of magnitude increase of the nonlocal perturbation to more than 8\%, due to a higher Fermi velocity $v_{F,\mathrm{Ti}} \simeq 1.79 \times 10^6$\,m/s and lower plasma and collision frequencies (see Supplementary information S3). Ti is also prone to oxidation, which can likely further alter its response. This comparison highlights that HIPs are not shielded from nonclassical effects, although still enable effective bypassing.

Furthermore, our findings in figure~\ref{fig5} demonstrate that HIPs enable the probing of interfacial layers between vdW crystals and screening materials, providing crucial insights into the optical properties and thickness of such layers. This approach is particularly effective when nonclassical effects are negligible, as in this study. However, Fig.~\ref{fig4} and \ref{fig5} also call attention to the fact that, while it is not yet common practice, all experiments investigating nonclassical effects should carefully examine the relevant interface, e.g. by using TEM.

Finally, our results suggest that HIPs serve as an excellent testbed for probing nonlocal effects in metals and assessing the accuracy of theoretical models describing them. In thin hBN films, HIPs can achieve extreme confinement levels where nonlocal metal models, such as the hydrodynamical model used here, remain untested. This raises a fundamental question: can our theoretical predictions be experimentally verified -- highlighting the surprising nature of virtually boundlessly confined polaritons -- or does the model break down, offering new insights into the microscopic behaviour of charge carriers in noble metals? Regardless of the outcome, these findings pave the way for exciting experiments that push polariton confinement further than ever before while addressing key open questions in the field.

\section{Methods}
\subsection{Sample preparation}
We synthesise monocrystalline Au flakes using a thermolysis-based Brust--Schiffrin method~\cite{Brust:1994}. An aqueous solution of chloroauric acid (HAuCl$_4$ $\cdot$ 3H$_2$O in a 5\,mM concentration) is mixed with tetraoctylammonium bromide in toluene, stirred for 10~minutes at 5000\,rpm, then allowed to separate into aqueous and organic phases over a further 10~minutes. All reagents were purchased from Sigma--Aldrich.

The Si substrate is prepared by precleaning in sequential ultrasonic baths: first, in acetone, followed by isopropyl alcohol (IPA), and finally with ultrapure water. After being blow-dried with nitrogen gas, the substrate is baked on a hot plate at $200^\circ\mathrm{C}$ for approximately 5~minutes to ensure dehydration. To induce the growth of Au flakes, a few microlitres of the organic phase is drop-cast onto the substrate, which is then left on a hot plate, enclosed under a beaker, at $130^\circ\mathrm{C}$ for 24~hours. After that, the sample is cleaned in toluene at $75^\circ\mathrm{C}$, followed by acetone and IPA, which effectively removes most of the residual organic solvents. Immediately before hBN transfer, we clean the Au flake using oxygen plasma at an O$_2$ flow rate of 30\,sccm and a power setting of 30\,W for 15~minutes (CUTE-1MPR-Dual mode from Femto Science Inc.) or 300\,W for 15~minutes (PE-50 from Plasma Etch Inc.) with similar outcome.

To transfer hBN to the Au flakes, we mechanically exfoliate epitaxial solidification-grown hBN, commercially sourced from 2D semiconductors USA, using low adhesion Nitto SPV224 tape (blue tape). We exfoliate the original hBN crystal until the tape is visibly covered, then switch to a fresh piece of tape for a last round of exfoliation on a polydimethylsiloxane (PDMS) stamp, commercially available from GELPAK.

We select the hBN flakes on the PDMS stamp by comparing their visual contrast, and the chosen flake is transferred onto the Au flake of choice using a manual transfer stage at a temperature of $80^\circ\mathrm{C}$.

\subsection{s-SNOM measurements}
We perform near-field imaging using a commercial s-SNOM (neaSNOM from attocube systems AG, formerly Neaspec) coupled to a tunable quantum cascade laser (MIRcat from Daylight Solutions). To probe the propagating phonon-polaritons, we employ Pt/Ir-coated AFM tips (ARROW-NCPt-50 from NanoWorld) operating in non-contact mode, with a tapping frequency of approximately 270\,kHz and an oscillation amplitude between 60 and 70\,nm. 

In s-SNOM, infrared radiation from the laser -- tuned to a frequency within the Reststrahlen band of hBN -- is focused onto the metallic AFM tip. The incident light scatters from both the tip and the edge of the hBN flake, facilitating the necessary momentum change to launch phonon-polaritons.

For image generation, we utilise the pseudo-heterodyne interferometric detection module, where background-free interferometric signals are obtained by demodulating the scattering amplitude (the output signal of the s-SNOM) at different $n$-order harmonics of the tapping frequency. We select the $n=3$ harmonic to ensure near-field images with minimal background noise -- the $n=4$ harmonic can also be employed for further background reduction, albeit with a lower signal-to-noise ratio.

Additionally, during the s-SNOM raster scans, of the surface, we simultaneously acquire topography data for the scanned area, thereby providing precise thickness measurements of the hBN flakes and ensuring consistency across measurements. Furthermore, we employ the AFM mode of the neaSNOM to measure the surface roughness of select Au flakes prior to hBN transfer.

\subsection{TEM}
We prepare cross-sectional TEM specimens using a focused ion beam (FIB; Thermo Fisher Scientific Helios NanoLab 450). To mitigate damage and amorphisation of the ion-beam-sensitive hBN crystal a fine milling process at 5\,kV is carefully performed, gradually reducing the sample thickness to approximately 50\,nm.

Cross-sectional high-resolution TEM and scanning TEM images of the interface between the hBN and Au crystals are acquired using a cold field emission gun (JEOL-ARM300F), equipped with a probe aberration corrector and operating at an accelerating voltage of 160\,keV. EELS measurements are conducted in dual mode to simultaneously collect zero-loss and core-loss spectra, minimising energy drift while measuring energy loss.

\subsection{Theoretical modelling}
The frequency- and momentum-dependent p-polarisation reflectance coefficient $r_\mathrm{p}(\omega,q)$ of a general heterostructure is calculated by solving Maxwell's equations within each layer under plane-wave illumination and subsequently finding the electric and magnetic fields and associated coefficients through the interlayer boundary conditions. Notably, we incorporate the nonlocal response of metallic layers following the hydrodynamical model~\cite{Moreau:2013,Raza:2015a,Dias:2018}, whereby nonlocal longitudinal electromagnetic modes are incorporated alongside local transverse ones. The detailed procedure, including all boundary conditions employed to solve the problem, is presented in the Supplementary information. 

To determine the dispersion relation of the polaritonic modes supported by the heterostructure, we solve the condition $1/r_\mathrm{p}(\omega,q)=0$, which, in our case, is done by setting a real frequency $\omega$ and numerically solving the equation to find the complex-valued momentum $q$ that satisfies it. From the resulting $(\omega,q)$ pairs, the polariton phase velocity is calculated as $v_\phi=\omega/\mathrm{Re}(q)$. The dispersion relation is finally fed into the purely polaritonic electric and magnetic fields in each material region (i.e. calculated in the absence of illumination source), from where the energy density distribution is subsequently calculated (see Supplementary information S1 for details).

We apply this method to two different HIP-supporting heterostructures: (1) Au/hBN/air and (2) Au/dielectric layer/hBN/air. In the latter case, the dielectric layer corresponds to the interfacial layer identified in the TEM images, with a thickness of $2$\,nm and refractive index of $1.2$ (see Fig.~\ref{fig3} and associated discussion). In both cases, the hBN layer thickness is varied, and its permittivity along the in-plane ($s=\parallel$) and out-of-plane ($s=\perp$) directions (with respect to its monolayers) is described through a Lorentzian model:
\begin{equation}
  \epsilon_{\mathrm{hBN},s}(\omega) = \epsilon_{\infty,s}\left( \frac{\omega_{\mathrm{LO},s}^2-\omega^2-{\rm i} \omega \Gamma_{s}}{\omega_{\mathrm{TO},s}^2-\omega^2-{\rm i} \omega \Gamma_{s}} \right),
\end{equation}
with the parameters shown in Table~S2 in the Supplementary information (see also the associated discussion for details). To model the local permittivity of the semi-infinite Au layer, we use the Drude model:
\begin{equation}\label{eq:Drude}
  \epsilon_\mathrm{Au}(\omega) = \epsilon_{b} - \frac{\omega_{p}^2}{\omega(\omega + {\rm i} \gamma_{m})},
\end{equation}
with the parameters $\epsilon_{b}=9.5$, $\hbar\omega_{p}=9.06$\,eV, and $\hbar\gamma_{m}=71$\,meV~\cite{Yu:2017}. Within the hydrodynamical model framework, nonlocality in Au is characterised by the parameter $\beta=\sqrt{3/5} v_F$~\cite{Raza:2015a,Wegner:2023,Dias:2018}, where $v_{F,\mathrm{Au}} \simeq 1.40 \times 10^6$\,m/s is the Fermi velocity of electrons in Au~\cite{AM1976}. 

Finally, we compare the HIPs in hBN/Au heterostructures with the AGPs supported in graphene/gap/Au ones. These are computed using the same formalism and describing graphene as an interfacial nonlocal surface conductivity modeled as described in Supplementary information section~S1. Throughout all figures, we used the Fermi level $E_F=0.5$\,eV and damping $\hbar\gamma=10$\,meV to describe the graphene layer.

\section{Data availability}

The data that underlie the findings of this study are available from the corresponding authors upon reasonable request.

\section{References}

\bibliographystyle{apsrev4-1}
\bibliography{references}

\begin{thebibliography}{52}%
\makeatletter
\providecommand \@ifxundefined [1]{%
 \@ifx{#1\undefined}
}%
\providecommand \@ifnum [1]{%
 \ifnum #1\expandafter \@firstoftwo
 \else \expandafter \@secondoftwo
 \fi
}%
\providecommand \@ifx [1]{%
 \ifx #1\expandafter \@firstoftwo
 \else \expandafter \@secondoftwo
 \fi
}%
\providecommand \natexlab [1]{#1}%
\providecommand \enquote  [1]{``#1''}%
\providecommand \bibnamefont  [1]{#1}%
\providecommand \bibfnamefont [1]{#1}%
\providecommand \citenamefont [1]{#1}%
\providecommand \href@noop [0]{\@secondoftwo}%
\providecommand \href [0]{\begingroup \@sanitize@url \@href}%
\providecommand \@href[1]{\@@startlink{#1}\@@href}%
\providecommand \@@href[1]{\endgroup#1\@@endlink}%
\providecommand \@sanitize@url [0]{\catcode `\\12\catcode `\$12\catcode `\&12\catcode `\#12\catcode `\^12\catcode `\_12\catcode `\%12\relax}%
\providecommand \@@startlink[1]{}%
\providecommand \@@endlink[0]{}%
\providecommand \url  [0]{\begingroup\@sanitize@url \@url }%
\providecommand \@url [1]{\endgroup\@href {#1}{\urlprefix }}%
\providecommand \urlprefix  [0]{URL }%
\providecommand \Eprint [0]{\href }%
\providecommand \doibase [0]{http://dx.doi.org/}%
\providecommand \selectlanguage [0]{\@gobble}%
\providecommand \bibinfo  [0]{\@secondoftwo}%
\providecommand \bibfield  [0]{\@secondoftwo}%
\providecommand \translation [1]{[#1]}%
\providecommand \BibitemOpen [0]{}%
\providecommand \bibitemStop [0]{}%
\providecommand \bibitemNoStop [0]{.\EOS\space}%
\providecommand \EOS [0]{\spacefactor3000\relax}%
\providecommand \BibitemShut  [1]{\csname bibitem#1\endcsname}%
\let\auto@bib@innerbib\@empty
\bibitem [{\citenamefont {Basov}\ \emph {et~al.}(2016)\citenamefont {Basov}, \citenamefont {Fogler},\ and\ \citenamefont {{Garc{\'i}a de Abajo}}}]{Basov:2016}%
  \BibitemOpen
  \bibfield  {author} {\bibinfo {author} {\bibfnamefont {D.~N.}\ \bibnamefont {Basov}}, \bibinfo {author} {\bibfnamefont {M.~M.}\ \bibnamefont {Fogler}}, \ and\ \bibinfo {author} {\bibfnamefont {F.~J.}\ \bibnamefont {{Garc{\'i}a de Abajo}}},\ }\href {\doibase 10.1126/science.aag1992} {\bibfield  {journal} {\bibinfo  {journal} {Science}\ }\textbf {\bibinfo {volume} {354}},\ \bibinfo {pages} {aag1992} (\bibinfo {year} {2016})}\BibitemShut {NoStop}%
\bibitem [{\citenamefont {Moon}\ \emph {et~al.}(2023)\citenamefont {Moon}, \citenamefont {Kim}, \citenamefont {Park}, \citenamefont {Im}, \citenamefont {Kim}, \citenamefont {Hwang},\ and\ \citenamefont {Kim}}]{Moon:2023}%
  \BibitemOpen
  \bibfield  {author} {\bibinfo {author} {\bibfnamefont {S.}~\bibnamefont {Moon}}, \bibinfo {author} {\bibfnamefont {J.}~\bibnamefont {Kim}}, \bibinfo {author} {\bibfnamefont {J.}~\bibnamefont {Park}}, \bibinfo {author} {\bibfnamefont {S.}~\bibnamefont {Im}}, \bibinfo {author} {\bibfnamefont {J.}~\bibnamefont {Kim}}, \bibinfo {author} {\bibfnamefont {I.}~\bibnamefont {Hwang}}, \ and\ \bibinfo {author} {\bibfnamefont {J.~K.}\ \bibnamefont {Kim}},\ }\href {\doibase 10.1002/adma.202204161} {\bibfield  {journal} {\bibinfo  {journal} {Advanced Materials}\ }\textbf {\bibinfo {volume} {35}},\ \bibinfo {pages} {2204161} (\bibinfo {year} {2023})}\BibitemShut {NoStop}%
\bibitem [{\citenamefont {Basov}\ \emph {et~al.}(2021)\citenamefont {Basov}, \citenamefont {Asenjo-Garcia}, \citenamefont {Schuck}, \citenamefont {Zhu},\ and\ \citenamefont {Rubio}}]{Basov:2021}%
  \BibitemOpen
  \bibfield  {author} {\bibinfo {author} {\bibfnamefont {D.~N.}\ \bibnamefont {Basov}}, \bibinfo {author} {\bibfnamefont {A.}~\bibnamefont {Asenjo-Garcia}}, \bibinfo {author} {\bibfnamefont {P.~J.}\ \bibnamefont {Schuck}}, \bibinfo {author} {\bibfnamefont {X.}~\bibnamefont {Zhu}}, \ and\ \bibinfo {author} {\bibfnamefont {A.}~\bibnamefont {Rubio}},\ }\href {\doibase doi:10.1515/nanoph-2020-0449} {\bibfield  {journal} {\bibinfo  {journal} {Nanophotonics}\ }\textbf {\bibinfo {volume} {10}},\ \bibinfo {pages} {549} (\bibinfo {year} {2021})}\BibitemShut {NoStop}%
\bibitem [{\citenamefont {Cirac\`{i}}\ \emph {et~al.}(2012)\citenamefont {Cirac\`{i}}, \citenamefont {Hill}, \citenamefont {Mock}, \citenamefont {Urzhumov}, \citenamefont {Fern\'{a}ndez-Dom\'{i}nguez}, \citenamefont {Maier}, \citenamefont {Pendry}, \citenamefont {Chilkoti},\ and\ \citenamefont {Smith}}]{Ciraci:2012}%
  \BibitemOpen
  \bibfield  {author} {\bibinfo {author} {\bibfnamefont {C.}~\bibnamefont {Cirac\`{i}}}, \bibinfo {author} {\bibfnamefont {R.~T.}\ \bibnamefont {Hill}}, \bibinfo {author} {\bibfnamefont {J.~J.}\ \bibnamefont {Mock}}, \bibinfo {author} {\bibfnamefont {Y.}~\bibnamefont {Urzhumov}}, \bibinfo {author} {\bibfnamefont {A.~I.}\ \bibnamefont {Fern\'{a}ndez-Dom\'{i}nguez}}, \bibinfo {author} {\bibfnamefont {S.~A.}\ \bibnamefont {Maier}}, \bibinfo {author} {\bibfnamefont {J.~B.}\ \bibnamefont {Pendry}}, \bibinfo {author} {\bibfnamefont {A.}~\bibnamefont {Chilkoti}}, \ and\ \bibinfo {author} {\bibfnamefont {D.~R.}\ \bibnamefont {Smith}},\ }\href {\doibase 10.1126/science.1224823} {\bibfield  {journal} {\bibinfo  {journal} {Science}\ }\textbf {\bibinfo {volume} {337}},\ \bibinfo {pages} {1072} (\bibinfo {year} {2012})}\BibitemShut {NoStop}%
\bibitem [{\citenamefont {Raza}\ \emph {et~al.}(2015{\natexlab{a}})\citenamefont {Raza}, \citenamefont {Kadkhodazadeh}, \citenamefont {Christensen}, \citenamefont {{Di Vece}}, \citenamefont {Wubs}, \citenamefont {Mortensen},\ and\ \citenamefont {Stenger}}]{Raza:2015b}%
  \BibitemOpen
  \bibfield  {author} {\bibinfo {author} {\bibfnamefont {S.}~\bibnamefont {Raza}}, \bibinfo {author} {\bibfnamefont {S.}~\bibnamefont {Kadkhodazadeh}}, \bibinfo {author} {\bibfnamefont {T.}~\bibnamefont {Christensen}}, \bibinfo {author} {\bibfnamefont {M.}~\bibnamefont {{Di Vece}}}, \bibinfo {author} {\bibfnamefont {M.}~\bibnamefont {Wubs}}, \bibinfo {author} {\bibfnamefont {N.~A.}\ \bibnamefont {Mortensen}}, \ and\ \bibinfo {author} {\bibfnamefont {N.}~\bibnamefont {Stenger}},\ }\href {\doibase 10.1038/ncomms9788} {\bibfield  {journal} {\bibinfo  {journal} {Nature Communications}\ }\textbf {\bibinfo {volume} {6}},\ \bibinfo {pages} {8788} (\bibinfo {year} {2015}{\natexlab{a}})}\BibitemShut {NoStop}%
\bibitem [{\citenamefont {Yang}\ \emph {et~al.}(2019)\citenamefont {Yang}, \citenamefont {Zhu}, \citenamefont {Yan}, \citenamefont {Agarwal}, \citenamefont {Zheng}, \citenamefont {Joannopoulos}, \citenamefont {Lalanne}, \citenamefont {Christensen}, \citenamefont {Berggren},\ and\ \citenamefont {Solja\v{c}i\'{c}}}]{Yang:2019}%
  \BibitemOpen
  \bibfield  {author} {\bibinfo {author} {\bibfnamefont {Y.}~\bibnamefont {Yang}}, \bibinfo {author} {\bibfnamefont {D.}~\bibnamefont {Zhu}}, \bibinfo {author} {\bibfnamefont {W.}~\bibnamefont {Yan}}, \bibinfo {author} {\bibfnamefont {A.}~\bibnamefont {Agarwal}}, \bibinfo {author} {\bibfnamefont {M.}~\bibnamefont {Zheng}}, \bibinfo {author} {\bibfnamefont {J.~D.}\ \bibnamefont {Joannopoulos}}, \bibinfo {author} {\bibfnamefont {P.}~\bibnamefont {Lalanne}}, \bibinfo {author} {\bibfnamefont {T.}~\bibnamefont {Christensen}}, \bibinfo {author} {\bibfnamefont {K.~K.}\ \bibnamefont {Berggren}}, \ and\ \bibinfo {author} {\bibfnamefont {M.}~\bibnamefont {Solja\v{c}i\'{c}}},\ }\href {\doibase 10.1038/s41586-019-1803-1} {\bibfield  {journal} {\bibinfo  {journal} {Nature}\ }\textbf {\bibinfo {volume} {576}},\ \bibinfo {pages} {248} (\bibinfo {year} {2019})}\BibitemShut {NoStop}%
\bibitem [{\citenamefont {Mortensen}(2021)}]{Mortensen:2021}%
  \BibitemOpen
  \bibfield  {author} {\bibinfo {author} {\bibfnamefont {N.~A.}\ \bibnamefont {Mortensen}},\ }\href {\doibase 10.1515/nanoph-2021-0156} {\bibfield  {journal} {\bibinfo  {journal} {Nanophotonics}\ }\textbf {\bibinfo {volume} {10}},\ \bibinfo {pages} {2563} (\bibinfo {year} {2021})}\BibitemShut {NoStop}%
\bibitem [{\citenamefont {Boroviks}\ \emph {et~al.}(2022)\citenamefont {Boroviks}, \citenamefont {Lin}, \citenamefont {Zenin}, \citenamefont {Ziegler}, \citenamefont {Dellith}, \citenamefont {Gon\c{c}alves}, \citenamefont {Wolff}, \citenamefont {Bozhevolnyi}, \citenamefont {Huang},\ and\ \citenamefont {Mortensen}}]{Boroviks:2022}%
  \BibitemOpen
  \bibfield  {author} {\bibinfo {author} {\bibfnamefont {S.}~\bibnamefont {Boroviks}}, \bibinfo {author} {\bibfnamefont {Z.-H.}\ \bibnamefont {Lin}}, \bibinfo {author} {\bibfnamefont {V.~A.}\ \bibnamefont {Zenin}}, \bibinfo {author} {\bibfnamefont {M.}~\bibnamefont {Ziegler}}, \bibinfo {author} {\bibfnamefont {A.}~\bibnamefont {Dellith}}, \bibinfo {author} {\bibfnamefont {P.~A.~D.}\ \bibnamefont {Gon\c{c}alves}}, \bibinfo {author} {\bibfnamefont {C.}~\bibnamefont {Wolff}}, \bibinfo {author} {\bibfnamefont {S.~I.}\ \bibnamefont {Bozhevolnyi}}, \bibinfo {author} {\bibfnamefont {J.-S.}\ \bibnamefont {Huang}}, \ and\ \bibinfo {author} {\bibfnamefont {N.~A.}\ \bibnamefont {Mortensen}},\ }\href {\doibase 10.1038/s41467-022-30737-2} {\bibfield  {journal} {\bibinfo  {journal} {Nature Communications}\ }\textbf {\bibinfo {volume} {13}},\ \bibinfo {pages} {3105} (\bibinfo {year} {2022})}\BibitemShut {NoStop}%
\bibitem [{\citenamefont {Zhu}\ \emph {et~al.}(2016)\citenamefont {Zhu}, \citenamefont {Esteban}, \citenamefont {Borisov}, \citenamefont {Baumberg}, \citenamefont {Nordlander}, \citenamefont {Lezec}, \citenamefont {Aizpurua},\ and\ \citenamefont {Crozier}}]{Zhu:2016}%
  \BibitemOpen
  \bibfield  {author} {\bibinfo {author} {\bibfnamefont {W.}~\bibnamefont {Zhu}}, \bibinfo {author} {\bibfnamefont {R.}~\bibnamefont {Esteban}}, \bibinfo {author} {\bibfnamefont {A.~G.}\ \bibnamefont {Borisov}}, \bibinfo {author} {\bibfnamefont {J.~J.}\ \bibnamefont {Baumberg}}, \bibinfo {author} {\bibfnamefont {P.}~\bibnamefont {Nordlander}}, \bibinfo {author} {\bibfnamefont {H.~J.}\ \bibnamefont {Lezec}}, \bibinfo {author} {\bibfnamefont {J.}~\bibnamefont {Aizpurua}}, \ and\ \bibinfo {author} {\bibfnamefont {K.~B.}\ \bibnamefont {Crozier}},\ }\href {\doibase 10.1038/ncomms11495} {\bibfield  {journal} {\bibinfo  {journal} {Nature Communications}\ }\textbf {\bibinfo {volume} {7}},\ \bibinfo {pages} {11495} (\bibinfo {year} {2016})}\BibitemShut {NoStop}%
\bibitem [{\citenamefont {Raza}\ \emph {et~al.}(2015{\natexlab{b}})\citenamefont {Raza}, \citenamefont {Bozhevolnyi}, \citenamefont {Wubs},\ and\ \citenamefont {Mortensen}}]{Raza:2015a}%
  \BibitemOpen
  \bibfield  {author} {\bibinfo {author} {\bibfnamefont {S.}~\bibnamefont {Raza}}, \bibinfo {author} {\bibfnamefont {S.~I.}\ \bibnamefont {Bozhevolnyi}}, \bibinfo {author} {\bibfnamefont {M.}~\bibnamefont {Wubs}}, \ and\ \bibinfo {author} {\bibfnamefont {N.~A.}\ \bibnamefont {Mortensen}},\ }\href {\doibase 10.1088/0953-8984/27/18/183204} {\bibfield  {journal} {\bibinfo  {journal} {Journal of Physics: Condensed Matter}\ }\textbf {\bibinfo {volume} {27}},\ \bibinfo {pages} {183204} (\bibinfo {year} {2015}{\natexlab{b}})}\BibitemShut {NoStop}%
\bibitem [{\citenamefont {Mortensen}\ \emph {et~al.}(2014)\citenamefont {Mortensen}, \citenamefont {Raza}, \citenamefont {Wubs}, \citenamefont {Søndergaard},\ and\ \citenamefont {Bozhevolnyi}}]{Mortensen:2014}%
  \BibitemOpen
  \bibfield  {author} {\bibinfo {author} {\bibfnamefont {N.~A.}\ \bibnamefont {Mortensen}}, \bibinfo {author} {\bibfnamefont {S.}~\bibnamefont {Raza}}, \bibinfo {author} {\bibfnamefont {M.}~\bibnamefont {Wubs}}, \bibinfo {author} {\bibfnamefont {T.}~\bibnamefont {Søndergaard}}, \ and\ \bibinfo {author} {\bibfnamefont {S.~I.}\ \bibnamefont {Bozhevolnyi}},\ }\href {\doibase 10.1038/ncomms4809} {\bibfield  {journal} {\bibinfo  {journal} {Nature Communications}\ }\textbf {\bibinfo {volume} {5}},\ \bibinfo {pages} {3809} (\bibinfo {year} {2014})}\BibitemShut {NoStop}%
\bibitem [{\citenamefont {Feibelman}(1982)}]{Feibelman:1982}%
  \BibitemOpen
  \bibfield  {author} {\bibinfo {author} {\bibfnamefont {P.~J.}\ \bibnamefont {Feibelman}},\ }\href {\doibase 10.1016/0079-6816(82)90001-6} {\bibfield  {journal} {\bibinfo  {journal} {Progess in Surface Science}\ }\textbf {\bibinfo {volume} {12}},\ \bibinfo {pages} {287} (\bibinfo {year} {1982})}\BibitemShut {NoStop}%
\bibitem [{\citenamefont {Yan}\ \emph {et~al.}(2015)\citenamefont {Yan}, \citenamefont {Wubs},\ and\ \citenamefont {Mortensen}}]{Yan:2015}%
  \BibitemOpen
  \bibfield  {author} {\bibinfo {author} {\bibfnamefont {W.}~\bibnamefont {Yan}}, \bibinfo {author} {\bibfnamefont {M.}~\bibnamefont {Wubs}}, \ and\ \bibinfo {author} {\bibfnamefont {N.~A.}\ \bibnamefont {Mortensen}},\ }\href {\doibase 10.1103/PhysRevLett.115.137403} {\bibfield  {journal} {\bibinfo  {journal} {Physical Review Letters}\ }\textbf {\bibinfo {volume} {115}},\ \bibinfo {pages} {137403} (\bibinfo {year} {2015})}\BibitemShut {NoStop}%
\bibitem [{\citenamefont {Christensen}\ \emph {et~al.}(2017)\citenamefont {Christensen}, \citenamefont {Yan}, \citenamefont {Jauho}, \citenamefont {Solja\v{c}i\'{c}},\ and\ \citenamefont {Mortensen}}]{Christensen:2017}%
  \BibitemOpen
  \bibfield  {author} {\bibinfo {author} {\bibfnamefont {T.}~\bibnamefont {Christensen}}, \bibinfo {author} {\bibfnamefont {W.}~\bibnamefont {Yan}}, \bibinfo {author} {\bibfnamefont {A.-P.}\ \bibnamefont {Jauho}}, \bibinfo {author} {\bibfnamefont {M.}~\bibnamefont {Solja\v{c}i\'{c}}}, \ and\ \bibinfo {author} {\bibfnamefont {N.~A.}\ \bibnamefont {Mortensen}},\ }\href {\doibase 10.1103/PhysRevLett.118.157402} {\bibfield  {journal} {\bibinfo  {journal} {Physical Review Letters}\ }\textbf {\bibinfo {volume} {118}},\ \bibinfo {pages} {157402} (\bibinfo {year} {2017})}\BibitemShut {NoStop}%
\bibitem [{\citenamefont {Alonso-González}\ \emph {et~al.}(2017)\citenamefont {Alonso-González}, \citenamefont {Nikitin}, \citenamefont {Gao}, \citenamefont {Woessner}, \citenamefont {Lundeberg}, \citenamefont {Principi}, \citenamefont {Forcellini}, \citenamefont {Yan}, \citenamefont {Vélez}, \citenamefont {Huber}, \citenamefont {Watanabe}, \citenamefont {Taniguchi}, \citenamefont {Casanova}, \citenamefont {Hueso}, \citenamefont {Polini}, \citenamefont {Hone}, \citenamefont {Koppens},\ and\ \citenamefont {Hillenbrand}}]{Alonso-Gonzalez:2017}%
  \BibitemOpen
  \bibfield  {author} {\bibinfo {author} {\bibfnamefont {P.}~\bibnamefont {Alonso-González}}, \bibinfo {author} {\bibfnamefont {A.~Y.}\ \bibnamefont {Nikitin}}, \bibinfo {author} {\bibfnamefont {Y.}~\bibnamefont {Gao}}, \bibinfo {author} {\bibfnamefont {A.}~\bibnamefont {Woessner}}, \bibinfo {author} {\bibfnamefont {M.~B.}\ \bibnamefont {Lundeberg}}, \bibinfo {author} {\bibfnamefont {A.}~\bibnamefont {Principi}}, \bibinfo {author} {\bibfnamefont {N.}~\bibnamefont {Forcellini}}, \bibinfo {author} {\bibfnamefont {W.}~\bibnamefont {Yan}}, \bibinfo {author} {\bibfnamefont {S.}~\bibnamefont {Vélez}}, \bibinfo {author} {\bibfnamefont {A.~J.}\ \bibnamefont {Huber}}, \bibinfo {author} {\bibfnamefont {K.}~\bibnamefont {Watanabe}}, \bibinfo {author} {\bibfnamefont {T.}~\bibnamefont {Taniguchi}}, \bibinfo {author} {\bibfnamefont {F.}~\bibnamefont {Casanova}}, \bibinfo {author} {\bibfnamefont {L.~E.}\ \bibnamefont {Hueso}}, \bibinfo {author} {\bibfnamefont {M.}~\bibnamefont {Polini}}, \bibinfo {author} {\bibfnamefont
  {J.}~\bibnamefont {Hone}}, \bibinfo {author} {\bibfnamefont {F.~H.~L.}\ \bibnamefont {Koppens}}, \ and\ \bibinfo {author} {\bibfnamefont {R.}~\bibnamefont {Hillenbrand}},\ }\href {\doibase 10.1038/nnano.2016.185} {\bibfield  {journal} {\bibinfo  {journal} {Nature Nanotechnology}\ }\textbf {\bibinfo {volume} {12}},\ \bibinfo {pages} {31} (\bibinfo {year} {2017})}\BibitemShut {NoStop}%
\bibitem [{\citenamefont {Menabde}\ \emph {et~al.}(2021)\citenamefont {Menabde}, \citenamefont {Lee}, \citenamefont {Lee}, \citenamefont {Ha}, \citenamefont {Heiden}, \citenamefont {Yoo}, \citenamefont {Kim}, \citenamefont {Low}, \citenamefont {Lee}, \citenamefont {Oh},\ and\ \citenamefont {Jang}}]{Menabde:2021}%
  \BibitemOpen
  \bibfield  {author} {\bibinfo {author} {\bibfnamefont {S.~G.}\ \bibnamefont {Menabde}}, \bibinfo {author} {\bibfnamefont {I.-H.}\ \bibnamefont {Lee}}, \bibinfo {author} {\bibfnamefont {S.}~\bibnamefont {Lee}}, \bibinfo {author} {\bibfnamefont {H.}~\bibnamefont {Ha}}, \bibinfo {author} {\bibfnamefont {J.~T.}\ \bibnamefont {Heiden}}, \bibinfo {author} {\bibfnamefont {D.}~\bibnamefont {Yoo}}, \bibinfo {author} {\bibfnamefont {T.-T.}\ \bibnamefont {Kim}}, \bibinfo {author} {\bibfnamefont {T.}~\bibnamefont {Low}}, \bibinfo {author} {\bibfnamefont {Y.~H.}\ \bibnamefont {Lee}}, \bibinfo {author} {\bibfnamefont {S.-H.}\ \bibnamefont {Oh}}, \ and\ \bibinfo {author} {\bibfnamefont {M.~S.}\ \bibnamefont {Jang}},\ }\href {\doibase 10.1038/s41467-021-21193-5} {\bibfield  {journal} {\bibinfo  {journal} {Nature Communications}\ }\textbf {\bibinfo {volume} {12}},\ \bibinfo {pages} {938} (\bibinfo {year} {2021})}\BibitemShut {NoStop}%
\bibitem [{\citenamefont {Lundeberg}\ \emph {et~al.}(2017)\citenamefont {Lundeberg}, \citenamefont {Gao}, \citenamefont {Asgari}, \citenamefont {Tan}, \citenamefont {Van~Duppen}, \citenamefont {Autore}, \citenamefont {Alonso-González}, \citenamefont {Woessner}, \citenamefont {Watanabe}, \citenamefont {Taniguchi}, \citenamefont {Hillenbrand}, \citenamefont {Hone}, \citenamefont {Polini},\ and\ \citenamefont {Koppens}}]{Lundeberg:2017}%
  \BibitemOpen
  \bibfield  {author} {\bibinfo {author} {\bibfnamefont {M.~B.}\ \bibnamefont {Lundeberg}}, \bibinfo {author} {\bibfnamefont {Y.}~\bibnamefont {Gao}}, \bibinfo {author} {\bibfnamefont {R.}~\bibnamefont {Asgari}}, \bibinfo {author} {\bibfnamefont {C.}~\bibnamefont {Tan}}, \bibinfo {author} {\bibfnamefont {B.}~\bibnamefont {Van~Duppen}}, \bibinfo {author} {\bibfnamefont {M.}~\bibnamefont {Autore}}, \bibinfo {author} {\bibfnamefont {P.}~\bibnamefont {Alonso-González}}, \bibinfo {author} {\bibfnamefont {A.}~\bibnamefont {Woessner}}, \bibinfo {author} {\bibfnamefont {K.}~\bibnamefont {Watanabe}}, \bibinfo {author} {\bibfnamefont {T.}~\bibnamefont {Taniguchi}}, \bibinfo {author} {\bibfnamefont {R.}~\bibnamefont {Hillenbrand}}, \bibinfo {author} {\bibfnamefont {J.}~\bibnamefont {Hone}}, \bibinfo {author} {\bibfnamefont {M.}~\bibnamefont {Polini}}, \ and\ \bibinfo {author} {\bibfnamefont {F.~H.~L.}\ \bibnamefont {Koppens}},\ }\href {\doibase 10.1126/science.aan2735} {\bibfield  {journal} {\bibinfo  {journal}
  {Science}\ }\textbf {\bibinfo {volume} {357}},\ \bibinfo {pages} {187} (\bibinfo {year} {2017})}\BibitemShut {NoStop}%
\bibitem [{\citenamefont {Gonçalves}\ \emph {et~al.}(2020)\citenamefont {Gonçalves}, \citenamefont {Stenger}, \citenamefont {Cox}, \citenamefont {Mortensen},\ and\ \citenamefont {Xiao}}]{Goncalves:2020}%
  \BibitemOpen
  \bibfield  {author} {\bibinfo {author} {\bibfnamefont {P.~A.~D.}\ \bibnamefont {Gonçalves}}, \bibinfo {author} {\bibfnamefont {N.}~\bibnamefont {Stenger}}, \bibinfo {author} {\bibfnamefont {J.~D.}\ \bibnamefont {Cox}}, \bibinfo {author} {\bibfnamefont {N.~A.}\ \bibnamefont {Mortensen}}, \ and\ \bibinfo {author} {\bibfnamefont {S.}~\bibnamefont {Xiao}},\ }\href {\doibase 10.1002/adom.201901473} {\bibfield  {journal} {\bibinfo  {journal} {Advanced Optical Materials}\ }\textbf {\bibinfo {volume} {8}},\ \bibinfo {pages} {1901473} (\bibinfo {year} {2020})}\BibitemShut {NoStop}%
\bibitem [{\citenamefont {Gon\c{c}alves}\ \emph {et~al.}(2021)\citenamefont {Gon\c{c}alves}, \citenamefont {Christensen}, \citenamefont {Peres}, \citenamefont {Jauho}, \citenamefont {Epstein}, \citenamefont {Koppens}, \citenamefont {Solja\v{c}i\'{c}},\ and\ \citenamefont {Mortensen}}]{Goncalves:2021}%
  \BibitemOpen
  \bibfield  {author} {\bibinfo {author} {\bibfnamefont {P.~A.~D.}\ \bibnamefont {Gon\c{c}alves}}, \bibinfo {author} {\bibfnamefont {T.}~\bibnamefont {Christensen}}, \bibinfo {author} {\bibfnamefont {N.~M.~R.}\ \bibnamefont {Peres}}, \bibinfo {author} {\bibfnamefont {A.-P.}\ \bibnamefont {Jauho}}, \bibinfo {author} {\bibfnamefont {I.}~\bibnamefont {Epstein}}, \bibinfo {author} {\bibfnamefont {F.~H.~L.}\ \bibnamefont {Koppens}}, \bibinfo {author} {\bibfnamefont {M.}~\bibnamefont {Solja\v{c}i\'{c}}}, \ and\ \bibinfo {author} {\bibfnamefont {N.~A.}\ \bibnamefont {Mortensen}},\ }\href {\doibase 10.1038/s41467-021-23061-8} {\bibfield  {journal} {\bibinfo  {journal} {Nature Communications}\ }\textbf {\bibinfo {volume} {12}},\ \bibinfo {pages} {3271} (\bibinfo {year} {2021})}\BibitemShut {NoStop}%
\bibitem [{\citenamefont {Berkowitz}\ \emph {et~al.}(2021)\citenamefont {Berkowitz}, \citenamefont {Kim}, \citenamefont {Ni}, \citenamefont {McLeod}, \citenamefont {Lo}, \citenamefont {Sun}, \citenamefont {Gu}, \citenamefont {Watanabe}, \citenamefont {Taniguchi}, \citenamefont {Millis}, \citenamefont {Hone}, \citenamefont {Fogler}, \citenamefont {Averitt},\ and\ \citenamefont {Basov}}]{Berkowitz:2021}%
  \BibitemOpen
  \bibfield  {author} {\bibinfo {author} {\bibfnamefont {M.~E.}\ \bibnamefont {Berkowitz}}, \bibinfo {author} {\bibfnamefont {B.~S.~Y.}\ \bibnamefont {Kim}}, \bibinfo {author} {\bibfnamefont {G.}~\bibnamefont {Ni}}, \bibinfo {author} {\bibfnamefont {A.~S.}\ \bibnamefont {McLeod}}, \bibinfo {author} {\bibfnamefont {C.~F.~B.}\ \bibnamefont {Lo}}, \bibinfo {author} {\bibfnamefont {Z.}~\bibnamefont {Sun}}, \bibinfo {author} {\bibfnamefont {G.}~\bibnamefont {Gu}}, \bibinfo {author} {\bibfnamefont {K.}~\bibnamefont {Watanabe}}, \bibinfo {author} {\bibfnamefont {T.}~\bibnamefont {Taniguchi}}, \bibinfo {author} {\bibfnamefont {A.~J.}\ \bibnamefont {Millis}}, \bibinfo {author} {\bibfnamefont {J.~C.}\ \bibnamefont {Hone}}, \bibinfo {author} {\bibfnamefont {M.~M.}\ \bibnamefont {Fogler}}, \bibinfo {author} {\bibfnamefont {R.~D.}\ \bibnamefont {Averitt}}, \ and\ \bibinfo {author} {\bibfnamefont {D.~N.}\ \bibnamefont {Basov}},\ }\href {\doibase 10.1021/acs.nanolett.0c03684} {\bibfield  {journal} {\bibinfo  {journal} {Nano
  Letters}\ }\textbf {\bibinfo {volume} {21}},\ \bibinfo {pages} {308} (\bibinfo {year} {2021})}\BibitemShut {NoStop}%
\bibitem [{\citenamefont {Costa}\ \emph {et~al.}(2021)\citenamefont {Costa}, \citenamefont {Gonçalves}, \citenamefont {Basov}, \citenamefont {Koppens}, \citenamefont {Mortensen},\ and\ \citenamefont {Peres}}]{Costa:2021}%
  \BibitemOpen
  \bibfield  {author} {\bibinfo {author} {\bibfnamefont {A.~T.}\ \bibnamefont {Costa}}, \bibinfo {author} {\bibfnamefont {P.~A.~D.}\ \bibnamefont {Gonçalves}}, \bibinfo {author} {\bibfnamefont {D.~N.}\ \bibnamefont {Basov}}, \bibinfo {author} {\bibfnamefont {F.~H.~L.}\ \bibnamefont {Koppens}}, \bibinfo {author} {\bibfnamefont {N.~A.}\ \bibnamefont {Mortensen}}, \ and\ \bibinfo {author} {\bibfnamefont {N.~M.~R.}\ \bibnamefont {Peres}},\ }\href {\doibase 10.1073/pnas.2012847118} {\bibfield  {journal} {\bibinfo  {journal} {Proceedings of the National Academy of Sciences}\ }\textbf {\bibinfo {volume} {118}},\ \bibinfo {pages} {e2012847118} (\bibinfo {year} {2021})}\BibitemShut {NoStop}%
\bibitem [{\citenamefont {Dias}\ \emph {et~al.}(2018)\citenamefont {Dias}, \citenamefont {Alcaraz~Iranzo}, \citenamefont {Gon\c{c}alves}, \citenamefont {Hajati}, \citenamefont {Bludov}, \citenamefont {Jauho}, \citenamefont {Mortensen}, \citenamefont {Koppens},\ and\ \citenamefont {Peres}}]{Dias:2018}%
  \BibitemOpen
  \bibfield  {author} {\bibinfo {author} {\bibfnamefont {E.~J.~C.}\ \bibnamefont {Dias}}, \bibinfo {author} {\bibfnamefont {D.}~\bibnamefont {Alcaraz~Iranzo}}, \bibinfo {author} {\bibfnamefont {P.~A.~D.}\ \bibnamefont {Gon\c{c}alves}}, \bibinfo {author} {\bibfnamefont {Y.}~\bibnamefont {Hajati}}, \bibinfo {author} {\bibfnamefont {Y.~V.}\ \bibnamefont {Bludov}}, \bibinfo {author} {\bibfnamefont {A.-P.}\ \bibnamefont {Jauho}}, \bibinfo {author} {\bibfnamefont {N.~A.}\ \bibnamefont {Mortensen}}, \bibinfo {author} {\bibfnamefont {F.~H.~L.}\ \bibnamefont {Koppens}}, \ and\ \bibinfo {author} {\bibfnamefont {N.~M.~R.}\ \bibnamefont {Peres}},\ }\href {\doibase 10.1103/PhysRevB.97.245405} {\bibfield  {journal} {\bibinfo  {journal} {Physical Review B}\ }\textbf {\bibinfo {volume} {97}},\ \bibinfo {pages} {245405} (\bibinfo {year} {2018})}\BibitemShut {NoStop}%
\bibitem [{\citenamefont {Dai}\ \emph {et~al.}(2014)\citenamefont {Dai}, \citenamefont {Fei}, \citenamefont {Ma}, \citenamefont {Rodin}, \citenamefont {Wagner}, \citenamefont {McLeod}, \citenamefont {Liu}, \citenamefont {Gannett}, \citenamefont {Regan}, \citenamefont {Watanabe}, \citenamefont {Taniguchi}, \citenamefont {Thiemens}, \citenamefont {Dominguez}, \citenamefont {Castro~Neto}, \citenamefont {Zettl}, \citenamefont {Keilmann}, \citenamefont {Jarillo-Herrero}, \citenamefont {Fogler},\ and\ \citenamefont {Basov}}]{Dai:2014}%
  \BibitemOpen
  \bibfield  {author} {\bibinfo {author} {\bibfnamefont {S.}~\bibnamefont {Dai}}, \bibinfo {author} {\bibfnamefont {Z.}~\bibnamefont {Fei}}, \bibinfo {author} {\bibfnamefont {Q.}~\bibnamefont {Ma}}, \bibinfo {author} {\bibfnamefont {A.~S.}\ \bibnamefont {Rodin}}, \bibinfo {author} {\bibfnamefont {M.}~\bibnamefont {Wagner}}, \bibinfo {author} {\bibfnamefont {A.~S.}\ \bibnamefont {McLeod}}, \bibinfo {author} {\bibfnamefont {M.~K.}\ \bibnamefont {Liu}}, \bibinfo {author} {\bibfnamefont {W.}~\bibnamefont {Gannett}}, \bibinfo {author} {\bibfnamefont {W.}~\bibnamefont {Regan}}, \bibinfo {author} {\bibfnamefont {K.}~\bibnamefont {Watanabe}}, \bibinfo {author} {\bibfnamefont {T.}~\bibnamefont {Taniguchi}}, \bibinfo {author} {\bibfnamefont {M.}~\bibnamefont {Thiemens}}, \bibinfo {author} {\bibfnamefont {G.}~\bibnamefont {Dominguez}}, \bibinfo {author} {\bibfnamefont {A.~H.}\ \bibnamefont {Castro~Neto}}, \bibinfo {author} {\bibfnamefont {A.}~\bibnamefont {Zettl}}, \bibinfo {author} {\bibfnamefont {F.}~\bibnamefont
  {Keilmann}}, \bibinfo {author} {\bibfnamefont {P.}~\bibnamefont {Jarillo-Herrero}}, \bibinfo {author} {\bibfnamefont {M.~M.}\ \bibnamefont {Fogler}}, \ and\ \bibinfo {author} {\bibfnamefont {D.~N.}\ \bibnamefont {Basov}},\ }\href {\doibase 10.1126/science.1246833} {\bibfield  {journal} {\bibinfo  {journal} {Science}\ }\textbf {\bibinfo {volume} {343}},\ \bibinfo {pages} {1125} (\bibinfo {year} {2014})}\BibitemShut {NoStop}%
\bibitem [{\citenamefont {Caldwell}\ \emph {et~al.}(2014)\citenamefont {Caldwell}, \citenamefont {Kretinin}, \citenamefont {Chen}, \citenamefont {Giannini}, \citenamefont {Fogler}, \citenamefont {Francescato}, \citenamefont {Ellis}, \citenamefont {Tischler}, \citenamefont {Woods}, \citenamefont {Giles}, \citenamefont {Hong}, \citenamefont {Watanabe}, \citenamefont {Taniguchi}, \citenamefont {Maier},\ and\ \citenamefont {Novoselov}}]{Caldwell:2014}%
  \BibitemOpen
  \bibfield  {author} {\bibinfo {author} {\bibfnamefont {J.~D.}\ \bibnamefont {Caldwell}}, \bibinfo {author} {\bibfnamefont {A.~V.}\ \bibnamefont {Kretinin}}, \bibinfo {author} {\bibfnamefont {Y.}~\bibnamefont {Chen}}, \bibinfo {author} {\bibfnamefont {V.}~\bibnamefont {Giannini}}, \bibinfo {author} {\bibfnamefont {M.~M.}\ \bibnamefont {Fogler}}, \bibinfo {author} {\bibfnamefont {Y.}~\bibnamefont {Francescato}}, \bibinfo {author} {\bibfnamefont {C.~T.}\ \bibnamefont {Ellis}}, \bibinfo {author} {\bibfnamefont {J.~G.}\ \bibnamefont {Tischler}}, \bibinfo {author} {\bibfnamefont {C.~R.}\ \bibnamefont {Woods}}, \bibinfo {author} {\bibfnamefont {A.~J.}\ \bibnamefont {Giles}}, \bibinfo {author} {\bibfnamefont {M.}~\bibnamefont {Hong}}, \bibinfo {author} {\bibfnamefont {K.}~\bibnamefont {Watanabe}}, \bibinfo {author} {\bibfnamefont {T.}~\bibnamefont {Taniguchi}}, \bibinfo {author} {\bibfnamefont {S.~A.}\ \bibnamefont {Maier}}, \ and\ \bibinfo {author} {\bibfnamefont {K.~S.}\ \bibnamefont {Novoselov}},\ }\href
  {\doibase 10.1038/ncomms6221} {\bibfield  {journal} {\bibinfo  {journal} {Nature Communications}\ }\textbf {\bibinfo {volume} {5}},\ \bibinfo {pages} {5221} (\bibinfo {year} {2014})}\BibitemShut {NoStop}%
\bibitem [{\citenamefont {Dai}\ \emph {et~al.}(2019)\citenamefont {Dai}, \citenamefont {Fang}, \citenamefont {Rivera}, \citenamefont {Stehle}, \citenamefont {Jiang}, \citenamefont {Shen}, \citenamefont {Tay}, \citenamefont {Ciccarino}, \citenamefont {Ma}, \citenamefont {Rodan-Legrain}, \citenamefont {Jarillo-Herrero}, \citenamefont {Teo}, \citenamefont {Fogler}, \citenamefont {Narang}, \citenamefont {Kong},\ and\ \citenamefont {Basov}}]{Dai:2019}%
  \BibitemOpen
  \bibfield  {author} {\bibinfo {author} {\bibfnamefont {S.}~\bibnamefont {Dai}}, \bibinfo {author} {\bibfnamefont {W.}~\bibnamefont {Fang}}, \bibinfo {author} {\bibfnamefont {N.}~\bibnamefont {Rivera}}, \bibinfo {author} {\bibfnamefont {Y.}~\bibnamefont {Stehle}}, \bibinfo {author} {\bibfnamefont {B.-Y.}\ \bibnamefont {Jiang}}, \bibinfo {author} {\bibfnamefont {J.}~\bibnamefont {Shen}}, \bibinfo {author} {\bibfnamefont {R.~Y.}\ \bibnamefont {Tay}}, \bibinfo {author} {\bibfnamefont {C.~J.}\ \bibnamefont {Ciccarino}}, \bibinfo {author} {\bibfnamefont {Q.}~\bibnamefont {Ma}}, \bibinfo {author} {\bibfnamefont {D.}~\bibnamefont {Rodan-Legrain}}, \bibinfo {author} {\bibfnamefont {P.}~\bibnamefont {Jarillo-Herrero}}, \bibinfo {author} {\bibfnamefont {E.~H.~T.}\ \bibnamefont {Teo}}, \bibinfo {author} {\bibfnamefont {M.~M.}\ \bibnamefont {Fogler}}, \bibinfo {author} {\bibfnamefont {P.}~\bibnamefont {Narang}}, \bibinfo {author} {\bibfnamefont {J.}~\bibnamefont {Kong}}, \ and\ \bibinfo {author} {\bibfnamefont {D.~N.}\
  \bibnamefont {Basov}},\ }\href {\doibase 10.1002/adma.201806603} {\bibfield  {journal} {\bibinfo  {journal} {Advanced Materials}\ }\textbf {\bibinfo {volume} {31}},\ \bibinfo {pages} {1806603} (\bibinfo {year} {2019})}\BibitemShut {NoStop}%
\bibitem [{\citenamefont {Ambrosio}\ \emph {et~al.}(2018)\citenamefont {Ambrosio}, \citenamefont {Tamagnone}, \citenamefont {Chaudhary}, \citenamefont {Jauregui}, \citenamefont {Kim}, \citenamefont {Wilson},\ and\ \citenamefont {Capasso}}]{Ambrosio:2018}%
  \BibitemOpen
  \bibfield  {author} {\bibinfo {author} {\bibfnamefont {A.}~\bibnamefont {Ambrosio}}, \bibinfo {author} {\bibfnamefont {M.}~\bibnamefont {Tamagnone}}, \bibinfo {author} {\bibfnamefont {K.}~\bibnamefont {Chaudhary}}, \bibinfo {author} {\bibfnamefont {L.~A.}\ \bibnamefont {Jauregui}}, \bibinfo {author} {\bibfnamefont {P.}~\bibnamefont {Kim}}, \bibinfo {author} {\bibfnamefont {W.~L.}\ \bibnamefont {Wilson}}, \ and\ \bibinfo {author} {\bibfnamefont {F.}~\bibnamefont {Capasso}},\ }\href {\doibase 10.1038/s41377-018-0039-4} {\bibfield  {journal} {\bibinfo  {journal} {Light: Science \& Applications}\ }\textbf {\bibinfo {volume} {7}},\ \bibinfo {pages} {27} (\bibinfo {year} {2018})}\BibitemShut {NoStop}%
\bibitem [{\citenamefont {Lee}\ \emph {et~al.}(2020)\citenamefont {Lee}, \citenamefont {He}, \citenamefont {Zhang}, \citenamefont {Liu}, \citenamefont {Edgar}, \citenamefont {Wang}, \citenamefont {Avouris}, \citenamefont {Low}, \citenamefont {Caldwell},\ and\ \citenamefont {Oh}}]{Lee:2020}%
  \BibitemOpen
  \bibfield  {author} {\bibinfo {author} {\bibfnamefont {I.-H.}\ \bibnamefont {Lee}}, \bibinfo {author} {\bibfnamefont {M.}~\bibnamefont {He}}, \bibinfo {author} {\bibfnamefont {X.}~\bibnamefont {Zhang}}, \bibinfo {author} {\bibfnamefont {S.}~\bibnamefont {Liu}}, \bibinfo {author} {\bibfnamefont {J.~H.}\ \bibnamefont {Edgar}}, \bibinfo {author} {\bibfnamefont {K.}~\bibnamefont {Wang}}, \bibinfo {author} {\bibfnamefont {P.}~\bibnamefont {Avouris}}, \bibinfo {author} {\bibfnamefont {T.}~\bibnamefont {Low}}, \bibinfo {author} {\bibfnamefont {J.~D.}\ \bibnamefont {Caldwell}}, \ and\ \bibinfo {author} {\bibfnamefont {S.-H.}\ \bibnamefont {Oh}},\ }\href {\doibase 10.1038/s41467-020-17424-w} {\bibfield  {journal} {\bibinfo  {journal} {Nature Communications}\ }\textbf {\bibinfo {volume} {11}},\ \bibinfo {pages} {3649} (\bibinfo {year} {2020})}\BibitemShut {NoStop}%
\bibitem [{\citenamefont {Menabde}\ \emph {et~al.}(2022{\natexlab{a}})\citenamefont {Menabde}, \citenamefont {Boroviks}, \citenamefont {Ahn}, \citenamefont {Heiden}, \citenamefont {Watanabe}, \citenamefont {Taniguchi}, \citenamefont {Low}, \citenamefont {Hwang}, \citenamefont {Mortensen},\ and\ \citenamefont {Jang}}]{Menabde:2022b}%
  \BibitemOpen
  \bibfield  {author} {\bibinfo {author} {\bibfnamefont {S.~G.}\ \bibnamefont {Menabde}}, \bibinfo {author} {\bibfnamefont {S.}~\bibnamefont {Boroviks}}, \bibinfo {author} {\bibfnamefont {J.}~\bibnamefont {Ahn}}, \bibinfo {author} {\bibfnamefont {J.~T.}\ \bibnamefont {Heiden}}, \bibinfo {author} {\bibfnamefont {K.}~\bibnamefont {Watanabe}}, \bibinfo {author} {\bibfnamefont {T.}~\bibnamefont {Taniguchi}}, \bibinfo {author} {\bibfnamefont {T.}~\bibnamefont {Low}}, \bibinfo {author} {\bibfnamefont {D.~K.}\ \bibnamefont {Hwang}}, \bibinfo {author} {\bibfnamefont {N.~A.}\ \bibnamefont {Mortensen}}, \ and\ \bibinfo {author} {\bibfnamefont {M.~S.}\ \bibnamefont {Jang}},\ }\href {\doibase 10.1126/sciadv.abn0627} {\bibfield  {journal} {\bibinfo  {journal} {Science Advances}\ }\textbf {\bibinfo {volume} {8}},\ \bibinfo {pages} {eabn0627} (\bibinfo {year} {2022}{\natexlab{a}})}\BibitemShut {NoStop}%
\bibitem [{\citenamefont {Menabde}\ \emph {et~al.}(2022{\natexlab{b}})\citenamefont {Menabde}, \citenamefont {Heiden}, \citenamefont {Cox}, \citenamefont {Mortensen},\ and\ \citenamefont {Jang}}]{Menabde:2022a}%
  \BibitemOpen
  \bibfield  {author} {\bibinfo {author} {\bibfnamefont {S.~G.}\ \bibnamefont {Menabde}}, \bibinfo {author} {\bibfnamefont {J.~T.}\ \bibnamefont {Heiden}}, \bibinfo {author} {\bibfnamefont {J.~D.}\ \bibnamefont {Cox}}, \bibinfo {author} {\bibfnamefont {N.~A.}\ \bibnamefont {Mortensen}}, \ and\ \bibinfo {author} {\bibfnamefont {M.~S.}\ \bibnamefont {Jang}},\ }\href {\doibase 10.1515/nanoph-2021-0693} {\bibfield  {journal} {\bibinfo  {journal} {Nanophotonics}\ }\textbf {\bibinfo {volume} {11}},\ \bibinfo {pages} {2433} (\bibinfo {year} {2022}{\natexlab{b}})}\BibitemShut {NoStop}%
\bibitem [{\citenamefont {Gubbin}\ and\ \citenamefont {De~Liberato}(2020)}]{Gubbin:2020}%
  \BibitemOpen
  \bibfield  {author} {\bibinfo {author} {\bibfnamefont {C.~R.}\ \bibnamefont {Gubbin}}\ and\ \bibinfo {author} {\bibfnamefont {S.}~\bibnamefont {De~Liberato}},\ }\href {\doibase 10.1103/PhysRevX.10.021027} {\bibfield  {journal} {\bibinfo  {journal} {Physical Review X}\ }\textbf {\bibinfo {volume} {10}},\ \bibinfo {pages} {021027} (\bibinfo {year} {2020})}\BibitemShut {NoStop}%
\bibitem [{\citenamefont {Rodríguez~Echarri}\ \emph {et~al.}(2021)\citenamefont {Rodríguez~Echarri}, \citenamefont {Gon\c{c}alves}, \citenamefont {Tserkezis}, \citenamefont {Garc{\'i}a~de Abajo}, \citenamefont {Mortensen},\ and\ \citenamefont {Cox}}]{RodriguezEcharri:2021}%
  \BibitemOpen
  \bibfield  {author} {\bibinfo {author} {\bibfnamefont {A.}~\bibnamefont {Rodríguez~Echarri}}, \bibinfo {author} {\bibfnamefont {P.~A.~D.}\ \bibnamefont {Gon\c{c}alves}}, \bibinfo {author} {\bibfnamefont {C.}~\bibnamefont {Tserkezis}}, \bibinfo {author} {\bibfnamefont {F.~J.}\ \bibnamefont {Garc{\'i}a~de Abajo}}, \bibinfo {author} {\bibfnamefont {N.~A.}\ \bibnamefont {Mortensen}}, \ and\ \bibinfo {author} {\bibfnamefont {J.~D.}\ \bibnamefont {Cox}},\ }\href {\doibase 10.1364/OPTICA.412122} {\bibfield  {journal} {\bibinfo  {journal} {Optica}\ }\textbf {\bibinfo {volume} {8}},\ \bibinfo {pages} {710} (\bibinfo {year} {2021})}\BibitemShut {NoStop}%
\bibitem [{\citenamefont {Jiménez-Riobóo}\ \emph {et~al.}(2018)\citenamefont {Jiménez-Riobóo}, \citenamefont {Artús}, \citenamefont {Cuscó}, \citenamefont {Taniguchi}, \citenamefont {Cassabois},\ and\ \citenamefont {Gil}}]{Jimenez-Rioboo:2018}%
  \BibitemOpen
  \bibfield  {author} {\bibinfo {author} {\bibfnamefont {R.~J.}\ \bibnamefont {Jiménez-Riobóo}}, \bibinfo {author} {\bibfnamefont {L.}~\bibnamefont {Artús}}, \bibinfo {author} {\bibfnamefont {R.}~\bibnamefont {Cuscó}}, \bibinfo {author} {\bibfnamefont {T.}~\bibnamefont {Taniguchi}}, \bibinfo {author} {\bibfnamefont {G.}~\bibnamefont {Cassabois}}, \ and\ \bibinfo {author} {\bibfnamefont {B.}~\bibnamefont {Gil}},\ }\href {\doibase 10.1063/1.5019629} {\bibfield  {journal} {\bibinfo  {journal} {Applied Physics Letters}\ }\textbf {\bibinfo {volume} {112}},\ \bibinfo {pages} {051905} (\bibinfo {year} {2018})}\BibitemShut {NoStop}%
\bibitem [{\citenamefont {Fei}\ \emph {et~al.}(2011)\citenamefont {Fei}, \citenamefont {Andreev}, \citenamefont {Bao}, \citenamefont {Zhang}, \citenamefont {McLeod}, \citenamefont {Wang}, \citenamefont {Stewart}, \citenamefont {Zhao}, \citenamefont {Dominguez}, \citenamefont {Thiemens}, \citenamefont {Fogler}, \citenamefont {Tauber}, \citenamefont {Castro-Neto}, \citenamefont {Lau}, \citenamefont {Keilmann},\ and\ \citenamefont {Basov}}]{Fei:2011}%
  \BibitemOpen
  \bibfield  {author} {\bibinfo {author} {\bibfnamefont {Z.}~\bibnamefont {Fei}}, \bibinfo {author} {\bibfnamefont {G.~O.}\ \bibnamefont {Andreev}}, \bibinfo {author} {\bibfnamefont {W.}~\bibnamefont {Bao}}, \bibinfo {author} {\bibfnamefont {L.~M.}\ \bibnamefont {Zhang}}, \bibinfo {author} {\bibfnamefont {A.~S.}\ \bibnamefont {McLeod}}, \bibinfo {author} {\bibfnamefont {C.}~\bibnamefont {Wang}}, \bibinfo {author} {\bibfnamefont {M.~K.}\ \bibnamefont {Stewart}}, \bibinfo {author} {\bibfnamefont {Z.}~\bibnamefont {Zhao}}, \bibinfo {author} {\bibfnamefont {G.}~\bibnamefont {Dominguez}}, \bibinfo {author} {\bibfnamefont {M.}~\bibnamefont {Thiemens}}, \bibinfo {author} {\bibfnamefont {M.~M.}\ \bibnamefont {Fogler}}, \bibinfo {author} {\bibfnamefont {M.~J.}\ \bibnamefont {Tauber}}, \bibinfo {author} {\bibfnamefont {A.~H.}\ \bibnamefont {Castro-Neto}}, \bibinfo {author} {\bibfnamefont {C.~N.}\ \bibnamefont {Lau}}, \bibinfo {author} {\bibfnamefont {F.}~\bibnamefont {Keilmann}}, \ and\ \bibinfo {author} {\bibfnamefont
  {D.~N.}\ \bibnamefont {Basov}},\ }\href {\doibase 10.1021/nl202362d} {\bibfield  {journal} {\bibinfo  {journal} {Nano Letters}\ }\textbf {\bibinfo {volume} {11}},\ \bibinfo {pages} {4701} (\bibinfo {year} {2011})}\BibitemShut {NoStop}%
\bibitem [{\citenamefont {Kowalski}\ \emph {et~al.}(2025)\citenamefont {Kowalski}, \citenamefont {Mueller}, \citenamefont {Álvarez Pérez}, \citenamefont {Obst}, \citenamefont {Diaz-Granados}, \citenamefont {Carini}, \citenamefont {Senarath}, \citenamefont {Dixit}, \citenamefont {Niemann}, \citenamefont {Iyer}, \citenamefont {Kaps}, \citenamefont {Wetzel}, \citenamefont {Klopf}, \citenamefont {Kravchenko}, \citenamefont {Wolf}, \citenamefont {Folland}, \citenamefont {Eng}, \citenamefont {Kehr}, \citenamefont {Alonso-Gonzales}, \citenamefont {Paarmann},\ and\ \citenamefont {Caldwell}}]{Paarmann:2024}%
  \BibitemOpen
  \bibfield  {author} {\bibinfo {author} {\bibfnamefont {R.~A.}\ \bibnamefont {Kowalski}}, \bibinfo {author} {\bibfnamefont {N.~S.}\ \bibnamefont {Mueller}}, \bibinfo {author} {\bibfnamefont {G.}~\bibnamefont {Álvarez Pérez}}, \bibinfo {author} {\bibfnamefont {M.}~\bibnamefont {Obst}}, \bibinfo {author} {\bibfnamefont {K.}~\bibnamefont {Diaz-Granados}}, \bibinfo {author} {\bibfnamefont {G.}~\bibnamefont {Carini}}, \bibinfo {author} {\bibfnamefont {A.}~\bibnamefont {Senarath}}, \bibinfo {author} {\bibfnamefont {S.}~\bibnamefont {Dixit}}, \bibinfo {author} {\bibfnamefont {R.}~\bibnamefont {Niemann}}, \bibinfo {author} {\bibfnamefont {R.~B.}\ \bibnamefont {Iyer}}, \bibinfo {author} {\bibfnamefont {F.~G.}\ \bibnamefont {Kaps}}, \bibinfo {author} {\bibfnamefont {J.}~\bibnamefont {Wetzel}}, \bibinfo {author} {\bibfnamefont {J.~M.}\ \bibnamefont {Klopf}}, \bibinfo {author} {\bibfnamefont {I.~I.}\ \bibnamefont {Kravchenko}}, \bibinfo {author} {\bibfnamefont {M.}~\bibnamefont {Wolf}}, \bibinfo {author}
  {\bibfnamefont {T.~G.}\ \bibnamefont {Folland}}, \bibinfo {author} {\bibfnamefont {L.~M.}\ \bibnamefont {Eng}}, \bibinfo {author} {\bibfnamefont {S.~C.}\ \bibnamefont {Kehr}}, \bibinfo {author} {\bibfnamefont {P.}~\bibnamefont {Alonso-Gonzales}}, \bibinfo {author} {\bibfnamefont {A.}~\bibnamefont {Paarmann}}, \ and\ \bibinfo {author} {\bibfnamefont {J.~D.}\ \bibnamefont {Caldwell}},\ }\href {\doibase 10.48550/arXiv.2502.09909} {\bibfield  {journal} {\bibinfo  {journal} {arXiv:2502.09909}\ } (\bibinfo {year} {2025}),\ 10.48550/arXiv.2502.09909}\BibitemShut {NoStop}%
\bibitem [{\citenamefont {Huber}\ \emph {et~al.}(2005)\citenamefont {Huber}, \citenamefont {Ocelic}, \citenamefont {Kazantsev},\ and\ \citenamefont {Hillenbrand}}]{Huber:2005}%
  \BibitemOpen
  \bibfield  {author} {\bibinfo {author} {\bibfnamefont {A.}~\bibnamefont {Huber}}, \bibinfo {author} {\bibfnamefont {N.}~\bibnamefont {Ocelic}}, \bibinfo {author} {\bibfnamefont {D.}~\bibnamefont {Kazantsev}}, \ and\ \bibinfo {author} {\bibfnamefont {R.}~\bibnamefont {Hillenbrand}},\ }\href {\doibase 10.1063/1.2032595} {\bibfield  {journal} {\bibinfo  {journal} {Applied Physics Letters}\ ,\ \bibinfo {pages} {081103}} (\bibinfo {year} {2005})}\BibitemShut {NoStop}%
\bibitem [{\citenamefont {Wong}\ \emph {et~al.}(2021)\citenamefont {Wong}, \citenamefont {Hu}, \citenamefont {W.}, \citenamefont {Guo}, \citenamefont {Fung}, \citenamefont {Zhu},\ and\ \citenamefont {Lau}}]{Wong:2021}%
  \BibitemOpen
  \bibfield  {author} {\bibinfo {author} {\bibfnamefont {K.~P.}\ \bibnamefont {Wong}}, \bibinfo {author} {\bibfnamefont {X.}~\bibnamefont {Hu}}, \bibinfo {author} {\bibfnamefont {L.~T.}\ \bibnamefont {W.}}, \bibinfo {author} {\bibfnamefont {X.}~\bibnamefont {Guo}}, \bibinfo {author} {\bibfnamefont {K.~H.}\ \bibnamefont {Fung}}, \bibinfo {author} {\bibfnamefont {Y.}~\bibnamefont {Zhu}}, \ and\ \bibinfo {author} {\bibfnamefont {S.~P.}\ \bibnamefont {Lau}},\ }\href {\doibase 10.1002/adom.202100294} {\bibfield  {journal} {\bibinfo  {journal} {Advanced Optical Materials}\ ,\ \bibinfo {pages} {2100294}} (\bibinfo {year} {2021})}\BibitemShut {NoStop}%
\bibitem [{\citenamefont {Casses}\ \emph {et~al.}(2022)\citenamefont {Casses}, \citenamefont {Kaltenecker}, \citenamefont {Xiao}, \citenamefont {Wubs},\ and\ \citenamefont {Stenger}}]{Casses:2022}%
  \BibitemOpen
  \bibfield  {author} {\bibinfo {author} {\bibfnamefont {L.~N.}\ \bibnamefont {Casses}}, \bibinfo {author} {\bibfnamefont {K.~J.}\ \bibnamefont {Kaltenecker}}, \bibinfo {author} {\bibfnamefont {S.}~\bibnamefont {Xiao}}, \bibinfo {author} {\bibfnamefont {M.}~\bibnamefont {Wubs}}, \ and\ \bibinfo {author} {\bibfnamefont {N.}~\bibnamefont {Stenger}},\ }\href {\doibase 10.1364/OE.454740} {\bibfield  {journal} {\bibinfo  {journal} {Optics Express}\ }\textbf {\bibinfo {volume} {30}},\ \bibinfo {pages} {11181} (\bibinfo {year} {2022})}\BibitemShut {NoStop}%
\bibitem [{\citenamefont {Jang}\ \emph {et~al.}(2024)\citenamefont {Jang}, \citenamefont {Menabde}, \citenamefont {Kiani}, \citenamefont {Heiden}, \citenamefont {Zenin}, \citenamefont {Mortensen}, \citenamefont {Tagliabue},\ and\ \citenamefont {Jang}}]{Jang:2024}%
  \BibitemOpen
  \bibfield  {author} {\bibinfo {author} {\bibfnamefont {M.}~\bibnamefont {Jang}}, \bibinfo {author} {\bibfnamefont {S.~G.}\ \bibnamefont {Menabde}}, \bibinfo {author} {\bibfnamefont {F.}~\bibnamefont {Kiani}}, \bibinfo {author} {\bibfnamefont {J.~T.}\ \bibnamefont {Heiden}}, \bibinfo {author} {\bibfnamefont {V.~A.}\ \bibnamefont {Zenin}}, \bibinfo {author} {\bibfnamefont {N.~A.}\ \bibnamefont {Mortensen}}, \bibinfo {author} {\bibfnamefont {G.}~\bibnamefont {Tagliabue}}, \ and\ \bibinfo {author} {\bibfnamefont {M.~S.}\ \bibnamefont {Jang}},\ }\href {\doibase doi:10.1103/PhysRevApplied.22.014076} {\bibfield  {journal} {\bibinfo  {journal} {Physical Review Applied}\ }\textbf {\bibinfo {volume} {22}},\ \bibinfo {pages} {014076} (\bibinfo {year} {2024})}\BibitemShut {NoStop}%
\bibitem [{\citenamefont {Casses}\ \emph {et~al.}(2024)\citenamefont {Casses}, \citenamefont {Zhou}, \citenamefont {Lin}, \citenamefont {Tan}, \citenamefont {Bendixen-Fernex~de Mongex}, \citenamefont {Kaltenecker}, \citenamefont {Xiao}, \citenamefont {Wubs},\ and\ \citenamefont {Stenger}}]{Casses:2024}%
  \BibitemOpen
  \bibfield  {author} {\bibinfo {author} {\bibfnamefont {L.~N.}\ \bibnamefont {Casses}}, \bibinfo {author} {\bibfnamefont {B.}~\bibnamefont {Zhou}}, \bibinfo {author} {\bibfnamefont {Q.}~\bibnamefont {Lin}}, \bibinfo {author} {\bibfnamefont {A.}~\bibnamefont {Tan}}, \bibinfo {author} {\bibfnamefont {D.-P.}\ \bibnamefont {Bendixen-Fernex~de Mongex}}, \bibinfo {author} {\bibfnamefont {K.~J.}\ \bibnamefont {Kaltenecker}}, \bibinfo {author} {\bibfnamefont {S.}~\bibnamefont {Xiao}}, \bibinfo {author} {\bibfnamefont {M.}~\bibnamefont {Wubs}}, \ and\ \bibinfo {author} {\bibfnamefont {N.}~\bibnamefont {Stenger}},\ }\href {\doibase 10.1021/acsphotonics.4c00580} {\bibfield  {journal} {\bibinfo  {journal} {ACS Photonics}\ }\textbf {\bibinfo {volume} {11}},\ \bibinfo {pages} {3593} (\bibinfo {year} {2024})}\BibitemShut {NoStop}%
\bibitem [{\citenamefont {Shi}\ \emph {et~al.}(2015)\citenamefont {Shi}, \citenamefont {Bechtel}, \citenamefont {Berweger}, \citenamefont {Sun}, \citenamefont {Zeng}, \citenamefont {Jin}, \citenamefont {Chang}, \citenamefont {Martin}, \citenamefont {Raschke},\ and\ \citenamefont {Wang}}]{Zhi:2015}%
  \BibitemOpen
  \bibfield  {author} {\bibinfo {author} {\bibfnamefont {Z.}~\bibnamefont {Shi}}, \bibinfo {author} {\bibfnamefont {H.~A.}\ \bibnamefont {Bechtel}}, \bibinfo {author} {\bibfnamefont {S.}~\bibnamefont {Berweger}}, \bibinfo {author} {\bibfnamefont {Y.}~\bibnamefont {Sun}}, \bibinfo {author} {\bibfnamefont {B.}~\bibnamefont {Zeng}}, \bibinfo {author} {\bibfnamefont {C.}~\bibnamefont {Jin}}, \bibinfo {author} {\bibfnamefont {H.}~\bibnamefont {Chang}}, \bibinfo {author} {\bibfnamefont {M.~C.}\ \bibnamefont {Martin}}, \bibinfo {author} {\bibfnamefont {M.~B.}\ \bibnamefont {Raschke}}, \ and\ \bibinfo {author} {\bibfnamefont {F.}~\bibnamefont {Wang}},\ }\href {\doibase 10.1021/acsphotonics.5b00007} {\bibfield  {journal} {\bibinfo  {journal} {Nano Letters}\ }\textbf {\bibinfo {volume} {2}},\ \bibinfo {pages} {790} (\bibinfo {year} {2015})}\BibitemShut {NoStop}%
\bibitem [{\citenamefont {Alcaraz~Iranzo}\ \emph {et~al.}(2018)\citenamefont {Alcaraz~Iranzo}, \citenamefont {Nanot}, \citenamefont {Dias}, \citenamefont {Epstein}, \citenamefont {Peng}, \citenamefont {Efetov}, \citenamefont {Lundeberg}, \citenamefont {Parret}, \citenamefont {Osmond}, \citenamefont {Hong}, \citenamefont {Kong}, \citenamefont {Englund}, \citenamefont {Peres},\ and\ \citenamefont {Koppens}}]{Iranzo:2018}%
  \BibitemOpen
  \bibfield  {author} {\bibinfo {author} {\bibfnamefont {D.}~\bibnamefont {Alcaraz~Iranzo}}, \bibinfo {author} {\bibfnamefont {S.}~\bibnamefont {Nanot}}, \bibinfo {author} {\bibfnamefont {E.~J.~C.}\ \bibnamefont {Dias}}, \bibinfo {author} {\bibfnamefont {I.}~\bibnamefont {Epstein}}, \bibinfo {author} {\bibfnamefont {C.}~\bibnamefont {Peng}}, \bibinfo {author} {\bibfnamefont {D.~K.}\ \bibnamefont {Efetov}}, \bibinfo {author} {\bibfnamefont {M.~B.}\ \bibnamefont {Lundeberg}}, \bibinfo {author} {\bibfnamefont {R.}~\bibnamefont {Parret}}, \bibinfo {author} {\bibfnamefont {J.}~\bibnamefont {Osmond}}, \bibinfo {author} {\bibfnamefont {J.-Y.}\ \bibnamefont {Hong}}, \bibinfo {author} {\bibfnamefont {J.}~\bibnamefont {Kong}}, \bibinfo {author} {\bibfnamefont {D.~R.}\ \bibnamefont {Englund}}, \bibinfo {author} {\bibfnamefont {N.~M.~R.}\ \bibnamefont {Peres}}, \ and\ \bibinfo {author} {\bibfnamefont {F.~H.~L.}\ \bibnamefont {Koppens}},\ }\href {\doibase 10.1126/science.aar8438} {\bibfield  {journal} {\bibinfo  {journal}
  {Science}\ }\textbf {\bibinfo {volume} {360}},\ \bibinfo {pages} {291} (\bibinfo {year} {2018})}\BibitemShut {NoStop}%
\bibitem [{\citenamefont {Smith}(1980)}]{Smith:1980}%
  \BibitemOpen
  \bibfield  {author} {\bibinfo {author} {\bibfnamefont {T.}~\bibnamefont {Smith}},\ }\href {\doibase 10.1016/0021-9797(80)90348-3} {\bibfield  {journal} {\bibinfo  {journal} {Journal of Colloid and Interface Science}\ }\textbf {\bibinfo {volume} {75}},\ \bibinfo {pages} {51} (\bibinfo {year} {1980})}\BibitemShut {NoStop}%
\bibitem [{\citenamefont {Turetta}\ \emph {et~al.}(2021)\citenamefont {Turetta}, \citenamefont {Sedona}, \citenamefont {Liscio}, \citenamefont {Sambi},\ and\ \citenamefont {Samori}}]{Turetta:2021}%
  \BibitemOpen
  \bibfield  {author} {\bibinfo {author} {\bibfnamefont {N.}~\bibnamefont {Turetta}}, \bibinfo {author} {\bibfnamefont {F.}~\bibnamefont {Sedona}}, \bibinfo {author} {\bibfnamefont {A.}~\bibnamefont {Liscio}}, \bibinfo {author} {\bibfnamefont {M.}~\bibnamefont {Sambi}}, \ and\ \bibinfo {author} {\bibfnamefont {P.}~\bibnamefont {Samori}},\ }\href {\doibase 10.1002/admi.202100068} {\bibfield  {journal} {\bibinfo  {journal} {Advanced Materials Interfaces}\ }\textbf {\bibinfo {volume} {8}},\ \bibinfo {pages} {2100068} (\bibinfo {year} {2021})}\BibitemShut {NoStop}%
\bibitem [{\citenamefont {Hong}\ \emph {et~al.}(2023)\citenamefont {Hong}, \citenamefont {Oh}, \citenamefont {Dat}, \citenamefont {Pak}, \citenamefont {Cha}, \citenamefont {Ko}, \citenamefont {Choi}, \citenamefont {Low}, \citenamefont {Oh},\ and\ \citenamefont {Kim}}]{Hong:2023}%
  \BibitemOpen
  \bibfield  {author} {\bibinfo {author} {\bibfnamefont {C.}~\bibnamefont {Hong}}, \bibinfo {author} {\bibfnamefont {S.}~\bibnamefont {Oh}}, \bibinfo {author} {\bibfnamefont {V.~K.}\ \bibnamefont {Dat}}, \bibinfo {author} {\bibfnamefont {S.}~\bibnamefont {Pak}}, \bibinfo {author} {\bibfnamefont {S.}~\bibnamefont {Cha}}, \bibinfo {author} {\bibfnamefont {K.-H.}\ \bibnamefont {Ko}}, \bibinfo {author} {\bibfnamefont {G.-M.}\ \bibnamefont {Choi}}, \bibinfo {author} {\bibfnamefont {T.}~\bibnamefont {Low}}, \bibinfo {author} {\bibfnamefont {S.-H.}\ \bibnamefont {Oh}}, \ and\ \bibinfo {author} {\bibfnamefont {J.-H.}\ \bibnamefont {Kim}},\ }\href {\doibase 10.1038/s41377-023-01308-x} {\bibfield  {journal} {\bibinfo  {journal} {Light: Science \& Applications}\ }\textbf {\bibinfo {volume} {12}},\ \bibinfo {pages} {280} (\bibinfo {year} {2023})}\BibitemShut {NoStop}%
\bibitem [{\citenamefont {Myers}\ \emph {et~al.}(2017)\citenamefont {Myers}, \citenamefont {Tonkyn}, \citenamefont {Danby}, \citenamefont {Taubman}, \citenamefont {Bernacki}, \citenamefont {Birnbaum}, \citenamefont {Sharpe},\ and\ \citenamefont {Johnson}}]{Myers:2017}%
  \BibitemOpen
  \bibfield  {author} {\bibinfo {author} {\bibfnamefont {T.~L.}\ \bibnamefont {Myers}}, \bibinfo {author} {\bibfnamefont {R.~G.}\ \bibnamefont {Tonkyn}}, \bibinfo {author} {\bibfnamefont {T.~O.}\ \bibnamefont {Danby}}, \bibinfo {author} {\bibfnamefont {M.~S.}\ \bibnamefont {Taubman}}, \bibinfo {author} {\bibfnamefont {B.~E.}\ \bibnamefont {Bernacki}}, \bibinfo {author} {\bibfnamefont {J.~C.}\ \bibnamefont {Birnbaum}}, \bibinfo {author} {\bibfnamefont {S.~W.}\ \bibnamefont {Sharpe}}, \ and\ \bibinfo {author} {\bibfnamefont {T.~J.}\ \bibnamefont {Johnson}},\ }\href {\doibase 10.1177/0003702817742848} {\bibfield  {journal} {\bibinfo  {journal} {Applied Spectroscopy}\ }\textbf {\bibinfo {volume} {72}},\ \bibinfo {pages} {535} (\bibinfo {year} {2017})}\BibitemShut {NoStop}%
\bibitem [{\citenamefont {Zhang}\ \emph {et~al.}(2020)\citenamefont {Zhang}, \citenamefont {Qiu}, \citenamefont {Zhao}, \citenamefont {Li},\ and\ \citenamefont {Liu}}]{Zhang:2020}%
  \BibitemOpen
  \bibfield  {author} {\bibinfo {author} {\bibfnamefont {X.}~\bibnamefont {Zhang}}, \bibinfo {author} {\bibfnamefont {J.}~\bibnamefont {Qiu}}, \bibinfo {author} {\bibfnamefont {J.}~\bibnamefont {Zhao}}, \bibinfo {author} {\bibfnamefont {X.}~\bibnamefont {Li}}, \ and\ \bibinfo {author} {\bibfnamefont {L.}~\bibnamefont {Liu}},\ }\href {\doibase 10.1016/j.jqsrt.2020.107063} {\bibfield  {journal} {\bibinfo  {journal} {Journal of Quantitative Spectroscopy \& Radiative Transfer}\ }\textbf {\bibinfo {volume} {252}},\ \bibinfo {pages} {107063} (\bibinfo {year} {2020})}\BibitemShut {NoStop}%
\bibitem [{\citenamefont {{F. Monticone, N. A. Mortensen, \emph{et al.}}}(2025)}]{Monticone:2025}%
  \BibitemOpen
  \bibfield  {author} {\bibinfo {author} {\bibnamefont {{F. Monticone, N. A. Mortensen, \emph{et al.}}}},\ }\href {\doibase 10.48550/arXiv.2503.00519} {\bibfield  {journal} {\bibinfo  {journal} {arXiv:2503.00519}\ } (\bibinfo {year} {2025}),\ 10.48550/arXiv.2503.00519}\BibitemShut {NoStop}%
\bibitem [{\citenamefont {Brust}\ \emph {et~al.}(1994)\citenamefont {Brust}, \citenamefont {Walker}, \citenamefont {Bethell}, \citenamefont {Schiffrin},\ and\ \citenamefont {Whyman}}]{Brust:1994}%
  \BibitemOpen
  \bibfield  {author} {\bibinfo {author} {\bibfnamefont {M.}~\bibnamefont {Brust}}, \bibinfo {author} {\bibfnamefont {M.}~\bibnamefont {Walker}}, \bibinfo {author} {\bibfnamefont {D.}~\bibnamefont {Bethell}}, \bibinfo {author} {\bibfnamefont {D.~J.}\ \bibnamefont {Schiffrin}}, \ and\ \bibinfo {author} {\bibfnamefont {R.}~\bibnamefont {Whyman}},\ }\href {\doibase 10.1039/C39940000801} {\bibfield  {journal} {\bibinfo  {journal} {Journal of the Chemical Society, Chemical Communications}\ }\textbf {\bibinfo {volume} {(7)}},\ \bibinfo {pages} {801} (\bibinfo {year} {1994})}\BibitemShut {NoStop}%
\bibitem [{\citenamefont {Moreau}\ \emph {et~al.}(2013)\citenamefont {Moreau}, \citenamefont {Cirac\`{\i}},\ and\ \citenamefont {Smith}}]{Moreau:2013}%
  \BibitemOpen
  \bibfield  {author} {\bibinfo {author} {\bibfnamefont {A.}~\bibnamefont {Moreau}}, \bibinfo {author} {\bibfnamefont {C.}~\bibnamefont {Cirac\`{\i}}}, \ and\ \bibinfo {author} {\bibfnamefont {D.~R.}\ \bibnamefont {Smith}},\ }\href {\doibase 10.1103/PhysRevB.87.045401} {\bibfield  {journal} {\bibinfo  {journal} {Physical Review B}\ }\textbf {\bibinfo {volume} {87}},\ \bibinfo {pages} {045401} (\bibinfo {year} {2013})}\BibitemShut {NoStop}%
\bibitem [{\citenamefont {Yu}\ \emph {et~al.}(2017)\citenamefont {Yu}, \citenamefont {Liz-Marz{\'a}n},\ and\ \citenamefont {Garc{\'\i}a~de Abajo}}]{Yu:2017}%
  \BibitemOpen
  \bibfield  {author} {\bibinfo {author} {\bibfnamefont {R.}~\bibnamefont {Yu}}, \bibinfo {author} {\bibfnamefont {L.~M.}\ \bibnamefont {Liz-Marz{\'a}n}}, \ and\ \bibinfo {author} {\bibfnamefont {F.~J.}\ \bibnamefont {Garc{\'\i}a~de Abajo}},\ }\href {\doibase 10.1039/C6CS00919K} {\bibfield  {journal} {\bibinfo  {journal} {Chemical Society Reviews}\ }\textbf {\bibinfo {volume} {46}},\ \bibinfo {pages} {6710} (\bibinfo {year} {2017})}\BibitemShut {NoStop}%
\bibitem [{\citenamefont {Wegner}\ \emph {et~al.}(2023)\citenamefont {Wegner}, \citenamefont {Huynh}, \citenamefont {Mortensen}, \citenamefont {Intravaia},\ and\ \citenamefont {Busch}}]{Wegner:2023}%
  \BibitemOpen
  \bibfield  {author} {\bibinfo {author} {\bibfnamefont {G.}~\bibnamefont {Wegner}}, \bibinfo {author} {\bibfnamefont {D.-N.}\ \bibnamefont {Huynh}}, \bibinfo {author} {\bibfnamefont {N.~A.}\ \bibnamefont {Mortensen}}, \bibinfo {author} {\bibfnamefont {F.}~\bibnamefont {Intravaia}}, \ and\ \bibinfo {author} {\bibfnamefont {K.}~\bibnamefont {Busch}},\ }\href {\doibase 10.1103/PhysRevB.107.115425} {\bibfield  {journal} {\bibinfo  {journal} {Physical Review B}\ }\textbf {\bibinfo {volume} {107}},\ \bibinfo {pages} {115425} (\bibinfo {year} {2023})}\BibitemShut {NoStop}%
\bibitem [{\citenamefont {Ashcroft}\ and\ \citenamefont {Mermin}(1976)}]{AM1976}%
  \BibitemOpen
  \bibfield  {author} {\bibinfo {author} {\bibfnamefont {N.~W.}\ \bibnamefont {Ashcroft}}\ and\ \bibinfo {author} {\bibfnamefont {N.~D.}\ \bibnamefont {Mermin}},\ }\href@noop {} {\emph {\bibinfo {title} {Solid State Physics}}}\ (\bibinfo  {publisher} {Harcourt College Publishers},\ \bibinfo {address} {Philadelphia},\ \bibinfo {year} {1976})\BibitemShut {NoStop}%
\end{thebibliography}%

\section{Acknowledgments} %
We acknowledge insightful discussions with C. Frydendahl. We thank N. Ubrig for reading and commenting on the manuscript.
This work was supported by the National Research Foundation of Korea (NRF) grants funded by the Korea government (MSIT) (RS-2022-NR070476, RS-2024-00412644, RS-2024-00332210, RS-2024-00340639, 2022R1A2C2011109).
The Center for Polariton-driven Light--Matter Interactions (POLIMA) is sponsored by the Danish National Research Foundation (Project No.~DNRF165).
V.~A.~Z. acknowledges financial support from VILLUM FONDEN (Grant No.~40707).

\section{Author Contributions Statement}

J.~T.~H., S.~G.~M., M.~S.~J., and N.~A.~M. conceived the overall idea.  
J.~T.~H., M.~N., and V.~A.~Z. processed and assembled the Au and hBN flakes, performing also the atomic-force microscopy measurements.
M.~K. and H.~Y.~H. performed the TEM imaging of samples.
J.~T.~H. performed the scanning near-field experiments.
J.~T.~H., E.~J.~C.~D., and N.~A.~M. contributed to the theoretical analysis.
All authors participated in the analysis of the data and the writing of the manuscript.

\vspace{5mm}
\section{Competing Interests Statement}
The authors declare no competing interests.

\end{document}